\documentclass[pre,twocolumn,floatfix,superscriptaddress,a4paper,nofootinbib]{revtex4-1}

\usepackage{graphicx}
\usepackage{color}
\usepackage{placeins}

\usepackage{hyperref}
\usepackage{amsmath}
\usepackage{amssymb}
\usepackage{mathtools}
\usepackage{bbold}
\usepackage{subfigure}

\usepackage[toc,page]{appendix}

\usepackage{siunitx}

\DeclareSIUnit\Molar{\textsc{M}}

\newcolumntype{K}{>{\centering\arraybackslash$}p{1cm}<{$}}

\def\epsilon{\varepsilon}
\def\theta{\vartheta}
\def\rho{\varrho}

\begin{document}

\title{Axisymmetric spheroidal squirmers and self-diffusiophoretic particles}

\date{\today}

\author{R. P\"ohnl}
\email{poehnl@hawaii.edu}
\affiliation{
	Max Planck Institute for Intelligent Systems, 
	Heisenbergstr.\ 3,
	70569 Stuttgart,
	Germany
}
\affiliation{
	IV$^{\text{th}}$ Institute for Theoretical Physics, 
	University of Stuttgart,
	Pfaffenwaldring 57,
	70569 Stuttgart,
	Germany
}
\affiliation{
Department of Mechanical Engineering,
University of Hawai'i at Manoa,
2540 Dole Street
Holmes 302
Honolulu, HI 96822,
USA
}

\author{M.N. Popescu}
\email{popescu@is.mpg.de}
\affiliation{
	Max Planck Institute for Intelligent Systems, 
	Heisenbergstr.\ 3,
	70569 Stuttgart,
	Germany
}
\author{W.E. Uspal}
\email{uspal@hawaii.edu}
\affiliation{
	Max Planck Institute for Intelligent Systems, 
	Heisenbergstr.\ 3,
	70569 Stuttgart,
	Germany
}
\affiliation{
	IV$^{\text{th}}$ Institute for Theoretical Physics, 
	University of Stuttgart,
	Pfaffenwaldring 57,
	70569 Stuttgart,
	Germany
}
\affiliation{
Department of Mechanical Engineering,
University of Hawai'i at Manoa,
2540 Dole Street
Holmes 302
Honolulu, HI 96822,
USA
}

\begin{abstract}
We study, by means of an exact analytical solution, the motion of a spheroidal, 
axisymmetric squirmer in an unbounded fluid, as well as the low Reynolds number 
hydrodynamic flow associated to it. In contrast to the case of a spherical 
squirmer --- for which, e.g., the velocity of the squirmer and the magnitude of 
the stresslet associated with the flow induced by the squirmer are respectively 
determined by the amplitudes of the first two slip (``squirming'') modes 
--- for the spheroidal squirmer each squirming mode either contributes to the 
velocity, or contributes to the stresslet. The results are straightforwardly 
extended to the self-phoresis of axisymmetric, spheroidal, chemically active 
particles in the case when the phoretic slip approximation holds.
\end{abstract}

\maketitle

\section{\label{Sec:Intro} Introduction}
Self-propulsion of micro-organisms, such as bacteria, algae and protozoa, plays 
an important role in many aspects of nature. Whether a bacteria tries to reach a 
nutrient rich area or a sperm cell an unfertilized egg, motility often yields a 
substantial advantage over competitors. Due to their small size and velocity, 
the viscous friction experienced by the micro-organisms swimming in 
water is very strong compared to inertial forces 
\cite{pedley1992hydrodynamic,guasto2012fluid}; consequently, they have developed 
swimming mechanisms adapted to these circumstances. 

The motion of motile bacteria and other small organisms is typically induced by the
beating of thin, thread-like appendages, the so called flagella; 
however, these species exhibit a broad morphology, in that they may posses 
either a single flagellum (e.g., monotrichous bacteria), several flagella 
(e.g., \textit{E. coli.}), or a carpet of many small flagella, called cilia 
(e.g. \textit{Opalina}). Focusing for the moment on the example of cilia 
covered micro-organisms, it was proposed by Lighthill that the emergence 
of motility can be understood without a detailed modeling of the complex, 
synchronous beating of the cilia \cite{lighthill1952squirming}. Instead, the 
effect of this beating pattern, i.e., the time-averaged surface flow induced by 
the envelope of cilia tips, has been modeled as providing a prescribed 
\textit{active} flow velocity $\mathbf{v}$ (actuation) at the surface of the 
particle (both within the tangent plane of the surface, $\mathbf{v}_s$, and in 
the direction normal to the surface, $\mathbf{v}_n$). The model, known by now as 
the ``squirmer'' model, was subsequently corrected and extended by his student 
Blake \cite{blake1971spherical}.

In its simplest and most used form, the tangential slip velocity of a spherical 
squirmer is taken to be the superposition of the fore-aft asymmetric 
and fore-aft symmetric modes with the slowest decaying contributions to flow in 
the surrounding liquid, i.e., the leading order modes. Combined, they make for 
a squirmer endowed with motility (from the leading fore-aft asymmetric 
mode) and with a hydrodynamic stresslet (from the leading fore-aft symmetric 
mode) 
\cite{zottl2014hydrodynamics,gotze2010mesoscale,zhu2012self,uspal2015rheotaxis,
wang2012inertial}.

Considerable progress has already been made in understanding the behavior of 
spherical squirmers, and many interesting questions have been answered, such as: 
what does the flow field around a squirmer look like and how it compares 
with the ones produced by simple microorganisms 
\cite{blake1971spherical,drescher2010direct,downton2009simulation}; what 
happens if 
the swimmer is not in free space, but rather disturbed by boundaries 
\cite{zottl2014hydrodynamics} or external flows \cite{uspal2015rheotaxis}; and 
how do pairs or even swarms of these particles interact 
\cite{gotze2010mesoscale}. Although this model can also be used to 
better understand the motion of micro-organisms, e.g. \textit{Volvox} 
\cite{pedley2016squirmers}, observed in experiments, its restriction to 
spherical swimmers limits a wider application. For example, \textit{Paramecia}, 
one of the most studied ciliates, has an elongated body 
\cite{sonneborn1970methods,zhang2015paramecia,ishikawa2006interaction}, which 
prevents a straightforward application of the traditional squirmer model. 
However, the aforementioned questions are also of interest for these and other 
elongated micro-organisms. A generalization of the squirmer model 
to non-spherical shapes is thus a natural, useful development.

Driven by advances in technology that allow increasingly sophisticated 
manufacturing capabilities, in the last decade significant efforts have
been made towards the development of artificial swimmers 
\cite{ismagilov2002autonomous,ozin2005dream,Faivre2017,ren2017rheotaxis,
golestanian2007designing}. The envisioned lab-on-a-chip devices and 
micro-cargo carriers \cite{ebbens2010pursuit,sundararajan2008catalytic}, e.g., 
for targeted drug deliveries and nanomachines focused on monitoring and 
dissolving harmful chemicals \cite{soler2014catalytic,gao2014environmental}, 
all need to perform precise motions on microscopic length scales. A better 
understanding of the general framework of microscale locomotion is required in 
order to optimally design and control such devices, and theoretical models 
facilitate new steps along this path. As one example, chemically active 
colloids achieve self-propulsion by harvesting local free energy. They catalyze 
a chemical reaction in the surrounding fluid and propel due to the ensuing 
chemical gradient. Spherical ``Janus'' particles belong to this group, and it 
has been shown that the squirmer model can, in various circumstances, capture 
essential features of their motion \cite{Popescu2018}. However, there are many 
chemically active colloids with non-spherical shape, and rod-like particles 
are especially prevalent in experimental studies (see, e.g., 
\cite{Paxton2004,Paxton2006, ren2017rheotaxis, mathijssen2018oscillatory}).

It is well known that passive, non-spherical colloids exposed to an ambient 
flow exhibit significant qualitative differences --- such as alignment with 
respect to the direction of the ambient flow \cite{uspal2013engineering}, 
Jeffery orbits by ellipsoids in shear flow \cite{jeffery1922motion}, nematic 
ordering arising from steric repulsion \cite{lettinga2005flow}, or 
noise-induced migration away from confining surfaces \cite{park2007cross} --- 
in comparison to their spherical counterparts. It is thus reasonable to expect 
that endowing such objects with a self-propulsion mechanism will lead to rich, 
qualitatively novel dynamical behaviors, some of which may be advantageous for, 
while others may hinder, applications. Accordingly, it is important to develop 
an in-depth understanding of the shape-dependent behavior, and significant 
experimental and theoretical efforts have been made in this direction (see, 
e.g., Refs. \cite{Kessler2004,Wensink2012,Goldstein2012, Frey2018,
Dogic2017,Goldstein2016,Clement2015,DiLeonardo2017,DiLeonardo2018,Poon2018}
as well as the insightful reviews provided in 
Refs. \cite{pedley1992hydrodynamic,Powers2009,Ebbens2010,Gompper2015_rev,Sen2010_rev,
Sagues_review_2018,Shelley2013,Ramaswamy2002,Bechinger2016_rev}).

Similar to the case of spherical swimmers, physical insight into the 
phenomenology exhibited by elongated swimmers can be gained from a corresponding 
squirmer model. For a model spheroidal swimmer moving by small deformations of 
its surface, Ref. \cite{felderhof2016stokesian} derived an analytic solution to 
the corresponding Stokes flow, from which the velocity of the swimmer could be 
calculated. In Ref. \cite{leshansky2007frictionless} it has been pointed out 
that an exact squirmer model for a spheroidal particle with a prescribed 
axi-symmetric, tangential slip velocity (active actuation of the fluid) on its 
surface can be written down by employing an available analytical solution
for the axi-symmetric Stokes flow around a spheroidal object 
\cite{dassios1994generalized}. However, the approach has been used in the 
context of a somewhat restricted model particle, involving an additional 
fore-aft asymmetry of the surface slip velocity, because of the particular 
interest, for that work, in the question of ``hydrodynamically stealthy'' 
microswimmers. While capturing the self-propulsion velocity of the swimmer, this 
removes certain characteristics of the flow, \textit{inter alia} those that 
contribute to the corresponding stresslet \cite{lauga2016stresslets} and would 
allow 
distinguishing between, e.g., ``puller'' and ``pusher'' type squirmers. 
Moreover, the particle stresslet is a key quantity connecting the microscopic 
dynamics of individual particles in a colloidal suspension with the continuum 
rheological properties of the suspension. For active particles like bacteria, the 
activity-induced stresslet can lead to novel material properties like "superfluidity" 
and spontaneous flow \cite{Clement2015}.

We note that also a strongly truncated model, based on an ansatz for the slip 
velocity in the form of two terms which, in the corresponding limit of a 
sphere, reproduce the first two modes of the spherical squirmer model of 
Lighthill and Blake, has been discussed by Ref. \cite{theers2016modeling}. 
This approach, however, is significantly affected by the fact that --- as 
noticed in Ref. \cite{felderhof2016stokesian} and also discussed here (see Sec. 
\ref{Sec:Squirmer}) --- for spheroidal squirmers both their velocity and their 
stresslet involve significant contributions from the higher order 
slip modes, in contrast to the case of a spherical squirmer for which 
only the first two modes contribute to those observables.

In this paper we employ the available analytical solution for axi-symmetric 
Stokes flow around a spheroidal object \cite{dassios1994generalized} to study 
the velocity and the induced hydrodynamic flow field around a spheroidal 
squirmer with a tangential slip velocity possessing axial symmetry, but 
otherwise unconstrained. The model squirmer is introduced in Sec. 
\ref{Sec:Model}. The series representation of the incompressible, 
axi-symmetric, creeping flow field around a spheroid 
\cite{dassios1994generalized} is succinctly summarized in 
Sec. \ref{Sec:Theory}. In Section \ref{Sec:Squirmer} we discuss the 
velocity and the flow field corresponding to the spheroidal squirmer, with 
particular emphasis on illustrating the contributions from the modes of various 
order. Additionally, in Sec. \ref{Sec:Phoretic} we discuss the straightforward 
extension of these 
results to deriving the flow field around a chemically active self-phoretic 
spheroid (for a similar mapping in the case of spherical particles see Ref. 
\cite{Michelin2014}). The final Sect. \ref{Sec:Summary} is devoted to a summary 
of the results and to the conclusions of the study.
 
\section{\label{Sec:Model} Model}

\begin{figure}[htb!]
\centering
 \subfigure[]{\includegraphics[width=0.5\columnwidth]{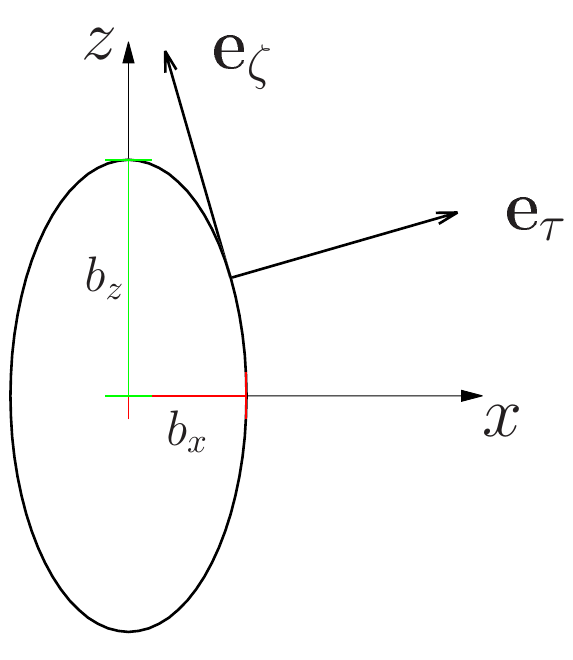}}
 \subfigure[]{\includegraphics[width=0.4\columnwidth]{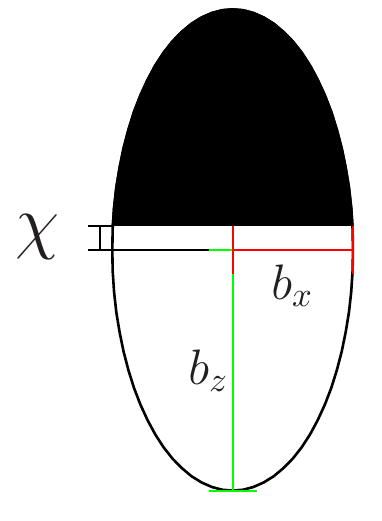}}
\caption{(a) Two-dimensional (2d) cut of a prolate spheroidal particle with semi-major 
axis $b_z$ and semi-minor axis $b_x$; also shown are the $x$ and $z$ axes of the 
co-moving system of coordinates and the unit vectors $\mathbf{e}_\tau$ and 
$\mathbf{e}_\zeta$ of the prolate spheroidal coordinate system; (b) 2d cut of a 
chemically-active self-phoretic particle; the chemically active part is the cap-like 
region shown in white, while the rest of the surface (black) is chemically inert. The 
quantity $\chi$ (``coverage'') denotes a certain measure for the 
extent of the active part (see Sec. \ref{Sec:Phoretic} in the main text).}
\label{models}
\end{figure}

The model system we consider is that of a spheroidal, rigid and impermeable particle 
immersed in an incompressible, unbounded, quiescent Newtonian liquid through which 
it moves due to a prescribed ``slip velocity'' (active actuation) at its surface 
(see Fig. \ref{models}). The slip velocity $\mathbf{v}_s$, which is tangential to the 
surface $\Sigma$ of the particle (i.e., $\mathbf{n} \cdot \mathbf{v}_s \equiv 0$ 
on $\Sigma$, with $\mathbf{n}$ denoting the outer (into the fluid) normal to $\Sigma$), 
is assumed to preserve the axial symmetry and to be constant in time, but it is 
otherwise arbitrary. The surface slip $\mathbf{v}_s$ is part of the model and thus it 
is a given function (or, alternatively as in, c.f., Sec. \ref{Sec:Phoretic}, it is 
determined as the solution of a separate problem). There are no external forces or 
torques acting on either the particle or the liquid. 

Due to the surface actuation, a hydrodynamic flow around the particle is induced 
and the particle sets in motion; we assume that the linear size $L$ of the 
particle, the viscosity $\mu$ and the density $\rho$ of the liquid, and the 
magnitude $|\mathbf{v}_s|$ of the slip velocity are such that the Reynolds 
number $\mathrm{Re} := \rho |\mathbf{v}_s| L/\mu$ is very small. Accordingly, 
after a short transient time a steady state hydrodynamic flow is induced around 
the particle and the particle translates steadily with velocity $\mathbf{U}$ 
(with respect to a fixed system of coordinates, the ``laboratory frame''). 
(Owing to the axial symmetry, in the absence of thermal fluctuations, which are 
neglected in this work, there is no rigid-body rotation of the particle in this 
model.) 

The analysis is more conveniently carried out in a system of coordinates 
attached to the particle (co-moving system). This is chosen with the origin at the 
center of the particle and such that the $z$-axis is along the axis of 
symmetry of the particle (thus $\mathbf{U} = U \mathbf{e}_z$). The 
semi-axis of the particle are accordingly denoted by $b_z$ 
and $b_x$ (see Fig. \ref{models}); their ratio $r_e = \frac{b_x}{b_z}$ is 
called the aspect ratio and determines the slenderness of the particle. The 
values $r_e < 1$, $r_e = 1$, and $r_e > 1$ correspond to prolate, spherical, 
and oblate shapes, respectively.

In the co-moving system of coordinates, the flow $\mathbf{v}$ and the velocity 
$\mathbf{U}$ are determined as the solution of the Stokes equations
\begin{equation}
\label{eq:Stokes}
 \nabla \cdot \boldsymbol{\sigma} = 0\,,~~\nabla \cdot \mathbf{v} = 0\,,
\end{equation}
where
\begin{equation}
 \label{eq:stress_tens}
 \boldsymbol{\sigma} := - p\, \mathcal{I} + \mu [\nabla \bold{v} 
 + (\nabla \bold{v})^T]\,,
\end{equation}
is the Newtonian stress tensor, with $p$ denoting the pressure (enforcing 
incompressibility), $\mathcal{I}$ denoting the unit tensor, and $()^T$ 
denoting the transpose, subject to: \newline
{--} the boundary conditions (BC)
\begin{subequations}
 \label{eq:BCs_Stokes}
 \begin{equation}
 \label{eq:BC_surf}
 \bold{v}|_\Sigma = \bold{v}_s\,,
 \end{equation}
and
\begin{equation}
\label{eq:BC_infty} 
\bold{v}(|\bold{r}| \to \infty) \rightarrow -\bold{U}\,.
\end{equation}
\end{subequations}
{--} the force balance condition (overdamped motion of the particle in the absence 
of external forces) 
\begin{equation}
 \label{eq:F_bal}
 \int\limits_\Sigma \boldsymbol{\sigma} \cdot \bold{n} \, d\Sigma = 0\,.
\end{equation}
The hydrodynamic flow $\bold{v}_\mathrm{lab}$ in the laboratory frame, if desired, is 
then obtained as $\bold{v}_\mathrm{lab}(\mathrm{P}) = \bold{v}(\mathrm{P}) + \bold{U}$, 
with $\mathrm{P}$ denoting the observation point.

The boundary-value problem above can be straightforwardly solved numerically by 
using, e.g., the Boundary Element Method (BEM) (see, e.g., 
Refs. 
\cite{ishimoto2017guidance,uspal2015rheotaxis,katuri2018cross,Simmchen2016}) 
as well as analytically. In the following, we shall focus on the analytical 
solution of this problem; the corresponding numerical solutions obtained by the 
BEM are presented in Appendix \ref{Sec:Num1} and used as a means of testing the 
convergence of the series representation of the analytical solution. Before 
proceeding, we note that we will focus the discussion on the case of prolate 
shapes, i.e., $r_e < 1$ (see Fig. \ref{models}(a)); the case of an oblate 
squirmer ($r_e > 1$) can be obtained from the results for the prolate shapes via 
a certain mapping (see, e.g., Ref. \cite{leshansky2007frictionless} and Appendix 
\ref{Sec:Oblate}), while the case of a sphere is obtained from the results 
corresponding to a prolate spheroid by taking the limit $r_e \to 1^{-}$.

\section{Hydrodynamic flow and velocity of a prolate squirmer\label{Sec:Theory}}

The calculation of the hydrodynamic flow $\bold{v}$ induced by the squirmer and 
of the velocity $\bold{U}$ of the squirmer are most conveniently carried out by 
employing the modified prolate spheroidal coordinates 
$(1\leq\tau\leq\infty,-1\leq\xi\leq1,0\leq\phi\leq2\pi)$ as in 
Ref. \cite{dassios1994generalized}. These are defined in the co-moving system of 
reference, which has the origin in the center of the particle, and are related 
to the Cartesian coordinates via \cite{dassios1994generalized}
\begin{align}
\label{eq:prol_coord}
\tau = \frac{1}{2c}(\sqrt{x^2+y^2+(z+c)^2}+ 
&\sqrt{x^2+y^2+(z-c)^2}\large{)}\,,\nonumber\\
\zeta= \frac{1}{2c}(\sqrt{x^2+y^2+(z+c)^2}-&\sqrt{x^2+y^2+(z-c)^2}),\\
\varphi = \arctan & \left(\frac{y}{x}\right)\,,\nonumber
\end{align}
where $c=\sqrt{b_z^2-b_x^2}$ is a purely geometric quantity which ensures smooth 
convergence into spherical coordinates in the limit $b_x \rightarrow b_z$. 
The unit vectors $\mathbf{e}_\tau$ and $\mathbf{e}_\zeta$ (see Fig. 
\ref{models}) are related to the ones of the Cartesian coordinates via
\begin{eqnarray}
 \label{eq:unit_vec}
\bold{e}_\tau &=& \left(\frac{2\tau \cdot \sqrt{1-\zeta^2}}{\sqrt{\tau^2-1}} 
\bold{e}_x + \zeta \bold{e}_z \right) \cdot 
\frac{\sqrt{\tau^2-1}}{\sqrt{\tau^2-\zeta^2}}\,, \nonumber\\
\bold{e}_\zeta &=& \left (\frac{2\zeta \cdot 
\sqrt{\tau^2-1}} {\sqrt{1-\zeta^2}} \bold{e}_x + 
\tau \bold{e}_z \right) \cdot \frac{\sqrt{1-\zeta^2}}{\sqrt{\tau^2-\zeta^2}}\,,
\end{eqnarray}
and the corresponding Lam{\'e} metric coefficients are given by
\begin{align}
\label{eq:metric_coef}
& h_\zeta =c\frac{\sqrt{\tau^2-\zeta^2}}{\sqrt{1-\zeta^2}}\,, \nonumber\\
& h_\tau=c\frac{\sqrt{\tau^2-\zeta^2}}{\sqrt{\tau^2-1}}\,, \\
& h_\varphi = c \sqrt{\tau^2-1}\sqrt{1-\zeta^2}\,.\nonumber
\end{align} 
The iso-surfaces $\tau = const$ are confocal prolate spheroids with common center 
$\mathrm{O}$; the surface of the particle corresponds to 
\begin{equation}
 \label{eq:tau_part}
 \tau_0 = \frac{b_z}{c} = \frac{1}{\sqrt{1-r_e^2}} > 1\,,
\end{equation}
and the values $\tau > \tau_0$ ($1 \leq \tau < \tau_0$) correspond to the 
exterior (interior) of the prolate ellipsoid, respectively. The coordinate 
$\zeta$ takes the values $\zeta = \pm 1$ at the $z = \pm b_z$ apexes, 
respectively, and $\zeta = 0$ on the equatorial ($x-y$ plane cut) circle. 
Note that, as shown in Fig. \ref{models}(a), $\bold{e}_\tau$ coincides with 
the normal $\mathbf{n}$ while $\mathbf{e}_\zeta$ and $\mathbf{e}_\phi$ span 
the tangent plane.

\subsection{\label{Sec:vel}Velocity of the squirmer}

The velocity of the squirmer can be determined from Eqs. (\ref{eq:Stokes}) 
- (\ref{eq:F_bal}) as a linear functional of the axi-symmetric slip velocity 
$\bold{v}_s = v_s(\zeta) \bold{e}_\zeta$ without explicitly solving for the 
flow $\bold{v}$. This follows via a straightforward application of the Lorentz 
reciprocal theorem \cite[Chapter 3-5]{happel2012low}. For the case of the 
prolate spheroid with $\bold{U}=U\bold{e}_z$, this renders (for details see, 
e.g., Refs. 
\cite{leshansky2007frictionless,popescu2010phoretic,lauga2016stresslets})
\begin{equation}
 U = -\frac{\tau_0}{2}\int\limits^1_{-1} \text{d}\zeta \,
 \sqrt{\frac{1-\zeta^2}{\tau_0^2-\zeta^2}} \, v_s(\zeta)\,.
 \label{eq:Ulorentz}
\end{equation} 
Therefore, $\bold{U}$ can be considered as known; accordingly, the 
boundary value problem defined by Eqs. (\ref{eq:Stokes}) - (\ref{eq:F_bal}) is 
specified and the solution $\bold{v}$ can be determined.

\subsection{\label{Sec:stream}Stokes stream function and hydrodynamic flow}
Since the boundary value problem defined by Eqs. (\ref{eq:Stokes}) - 
(\ref{eq:BCs_Stokes}) has axial symmetry, one searches for an axisymmetric 
solution $\bold{v}(\bold{r})$ expressed in terms of a Stokes stream function 
$\psi(\bold{r})$ as $\bold{v}(\bold{r}) = \nabla \times \frac{\psi(\bold{r})}{h_\varphi}\bold{e}_\varphi$ 
\cite[Chapter 4]{happel2012low}; this renders
\begin{align}
 v_\tau(\tau,\zeta) 
 &= \frac{1}{h_\zeta h_\varphi}\frac{\partial \psi}{\partial 
\zeta}\,,\label{velos}\\
v_\zeta(\tau,\zeta) &= - \frac{1}{h_\tau h_\varphi}\frac{\partial \psi}{\partial \tau}.
\label{velo1}
\end{align}
The general ``semiseparable'' solution for the stream function in prolate coordinates 
has been derived by Ref. \cite{dassios1994generalized} in the form of a series representation
\begin{align}
 \psi(\tau,\zeta)& = 
g_0(\tau)G_0(\zeta)+g_1(\tau)G_1(\zeta) \nonumber\\
 &+\sum_{n=2}^{\infty}[g_n(\tau)G_n(\zeta)+h_n(\tau)H_n(\zeta)].
 \label{Ansatz1}
\end{align}
where $G_n$ and $H_n$ are the Gegenbauer functions of the first and second kind, 
respectively (see Ref. \cite{Abramowitz}). The functions $g_n(\tau)$ and 
$h_n(\tau)$ are given by certain linear combinations of $G_k(\tau)$ and 
$H_k(\tau)$, the coefficients of which are fixed by the corresponding boundary 
conditions (see below). 

The general solution above is applied to our particular system as 
follows. Noting that for our system the solution $\bold{v}(\bold{r})$ should be 
bounded (i.e., $|\bold{v}(\bold{r})| < \infty$ everywhere), none of the terms 
involving the Gegenbauer functions of the second kind $H_n(\zeta)$, which are 
divergent along the $z$-axis (i.e., $\zeta = \pm 1$) \cite[Chapter 
4-23]{happel2012low}, can be present in the solution; accordingly, 
$h_n(\tau) \equiv 0$ 
for all $n \geq 2$. Furthermore, due to the same requirement of bounded 
magnitude of the flow, the terms involving the Gegenbauer functions of the 
first kind $G_0(\zeta) = 1$ and $G_1(\zeta) = - \zeta$ also cannot be present in 
the solution because they lead to divergences of $v_\zeta$ at $\zeta = \pm 1$ 
(see Eqs. (\ref{velo1}) and (\ref{eq:metric_coef})); accordingly, $g_0(\tau) 
\equiv 0$ and $g_1(\tau) \equiv 0$. Therefore, for our system only the functions 
$g_{n \geq 2}$ will be of interest; these are given by 
\cite{dassios1994generalized} 
\begin{align}
 g_2(\tau)=&\frac{C_2}{6}G_1(\tau)+C_4H_4(\tau)+D_2H_2(\tau)\nonumber\\
 &+F_2 G_2(\tau)+E_4 G_4(\tau)\nonumber\\
 g_3(\tau)=&-\frac{C_3}{90}G_0(\tau)+C_5H_5(\tau)+D_3H_3(\tau)\label{coefficients}
\\\nonumber
 &+F_3 G_3(\tau)+E_5 G_5(\tau) \\
 g_{n\ge 4 }(\tau)=& C_{n+2}H_{n+2}(\tau) + C_{n}H_{n-2}(\tau)+D_nH_n(\tau)\nonumber\\
 &+F_n G_n(\tau)+E_{n+2} G_{n+2}(\tau)+E_{n} G_{n-2}(\tau),\nonumber
 \end{align}
where the constants $\{C_n, D_n\}_{n \geq 2}$, $\{E_n\}_{n \geq 4}$, and 
$\{F_n\}_{n \geq 2}$, are fixed, as noted above, by 
requiring that the solution satisfies the BCs, Eqs. (\ref{eq:BCs_Stokes}), 
and the force balance condition, Eq. (\ref{eq:F_bal}).

Since the hydrodynamic force is proportional to the coefficient $C_2$ (see 
$g_2(\tau)$) \cite{leshansky2007frictionless}, the force balance (Eq. 
(\ref{eq:F_bal})) implies
\begin{subequations}
\label{eq:cond_coef_gn}
\begin{equation}
 \label{eq:forc_bal_psi}
 C_2 = 0\,.
\end{equation}
The boundary condition at infinity (Eq. (\ref{eq:BC_infty})) implies the asymptotic 
behavior of the stream function
\begin{equation}
 \psi(\tau\to \infty,\zeta) \propto \frac{1}{2} U c^2 (\tau^2-1)(1-\zeta^2)\,; 
 \nonumber
\end{equation}
by noting the asymptotic behaviors $G_n(\tau \gg 1) \propto \tau^n$ and 
$H_n(\tau \gg 1) \propto \tau^{-(n+1)}$ \cite{Abramowitz}, this implies 
that the functions $g_{n\geq 2}(\tau)$ cannot have contributions from terms 
involving the Gegenbauer functions $G_n(\tau)$ of index $n > 2$, i.e.,
\begin{equation}
 \label{eq:far_field_psi_1}
 E_n = 0\,,~n = 4,5,\dots~~\mathrm{and}~~F_n = 0\,,n = 3,4,\dots 
\end{equation}
Furthermore, in order to exactly match the asymptotic $(\tau^2-1)(1-\zeta^2)$ 
form and its prefactor, it is necessary that 
\begin{equation}
F_2 = -2 c^2 U\,.
\label{eq:velo_bc}
\end{equation}
The impenetrability of the surface and Eq. (\ref{velos}) imply that 
$\psi(\tau_0,\zeta)$ is a constant; since $\psi(\tau_0,\pm 1) = 0$ as 
$G_n(\zeta=\pm 1)=0$, one concludes that $\psi(\tau_0,\zeta) \equiv 0$ and thus
\label{eq:system_for_gns}
\begin{equation}
 \label{eq:BC_norm_psi}
 g_n(\tau_0) = 0,~n = 2,3,\dots
\end{equation}
Finally, the tangential slip velocity condition at the surface of the particle 
(Eq. (\ref{eq:BC_surf})) together with: the expression for $v_\zeta$ in Eq. 
(\ref{velo1}), the orthogonality of the Gegenbauer functions $G_{n\geq 
2}(\zeta)$ (see Appendix \ref{Sec:Ortho}), and the relation between the 
Gegenbauer functions $G_{n\geq 2}(\zeta)$ and the associated Legendre 
polynomials $P_{n\geq 1}^1(\zeta)$ (see the Appendix \ref{Sec:Ident}) 
leads to
\begin{align}
 \label{eq:BC_tang_psi}
 \left.\frac{d g_n}{d\tau}\right|_{\tau = \tau_0} =& 
c^2(n-\frac{1}{2})
 \int\limits_{-1}^{+1} \mathrm{d} \zeta (\tau_0^2-\zeta^2)^{1/2} v_s(\zeta) 
 P_{n-1}^1(\zeta)\,,\nonumber\\
 &~n = 2,3,\dots. 
\end{align}
\end{subequations}
For given $v_s(\zeta)$, and with $U$ determined by Eq. (\ref{eq:Ulorentz}), 
Eqs. (\ref{eq:BC_norm_psi}) and (\ref{eq:BC_tang_psi}) provide a system 
of linear equations determining all the remaining unknown coefficients 
$\{C_{n \geq 3},D_{n \geq 2}\}$, as detailed in the Appendix 
\ref{Sec:Coef}.

\section{Squirmer and squirming modes\label{Sec:Squirmer}}
The form of Eq. (\ref{eq:BC_tang_psi}) suggests (for the expansion of the slip velocity $\mathbf{v}_s$) the functions 
\begin{equation}
\label{eq:def_Vn}
 V_n(\zeta) := (\tau_0^2-\zeta^2)^{-1/2} P_n^1(\zeta)\,,
\end{equation}
as a suitable basis over the space of square integrable functions $f(\zeta)$ 
satisfying $f(\zeta = \pm 1) = 0$ (note that $V_n$ has a parametric dependence 
on $\tau_0 > 1$). Defining, in this space, the weighted scalar product
\begin{equation}
 \label{eq:scalar_prod}
 \langle f_1(\zeta), f_2(\zeta) \rangle_{w(\zeta)} :=
 \int\limits_{-1}^{+1} \mathrm{d} \zeta w(\zeta) f_1(\zeta) f_2(\zeta)\,,
\end{equation}
and choosing the weight as 
\begin{equation}
 \label{eq:def_weight}
 w(\zeta) = \tau_0^2-\zeta^2 > 0\,,
\end{equation}
one infers (form the known properties of the associated Legendre polynomials) 
that indeed the set $\{V_n\}_{n \geq 1}$ is an orthogonal basis,
\begin{equation}
 \label{eq:orthog}
 \langle V_n(\zeta), V_m(\zeta) \rangle_{w(\zeta)} = 
 \frac{n (n+1)}{n+1/2} \delta_{n,m}\,.
\end{equation}
Accordingly, the slip velocity function $\bold{v}_s = v_s(\zeta) 
\bold{e}_\zeta$ has a unique representation in terms of a series in the 
functions $\{V_n\}_{n \geq 1}$, 
\begin{equation}
\label{squirmmodel}
 v_s(\zeta) = \tau_0 \sum\limits_{n \geq 1} B_n V_n(\zeta)\,.
\end{equation}
The coefficients of the expansion are written in the form above to ensure 
that in the limit of a sphere ($b_x \to b_z$, i.e, $\tau_0 \to \infty$) one 
arrives at the usual form employed for a spherical squirmer, i.e., that of an 
expansion in associated Legendre polynomials $P_n^1(\cos\theta)$ (see, e.g., 
Ref. \cite{blake1971spherical,ishikawa2008coherent}). This can be seen by 
noticing that in the limit $\tau_0 \to \infty$ one has $\tau_0 
(\tau_0^2-\zeta^2)^{-1/2} 
P_n^1(\zeta) \to P_n^1(\zeta)$ and that $b_x \to b_z$ implies $\zeta \to 
\cos(\theta)$; accordingly, it follows that, in the limit of the 
shape approaching that of a sphere, $\tau_0 V_n(\zeta) \mathbf{e}_\zeta \to - 
P_n^1(\cos\theta) \mathbf{e}_\theta$ and the expansion in Ref. 
\cite{blake1971spherical} is matched identically upon changing $B_n \rightarrow \frac{2}{n(n+1)} B_n$.

With the representation of the slip velocity in terms of the function $V_n$, Eq. 
\eqref{squirmmodel}, upon exploiting the orthogonality relation (Eq. 
(\ref{eq:orthog})) the boundary condition in Eq. \eqref{eq:BC_tang_psi} renders 
the simple relations
 \begin{align}
 \left.\frac{\partial 
g_{n}(\tau)}{\partial\tau}\right|_{\tau=\tau_0}&=\tau_0 c^2 
n (n-1)B_{n-1} \,,~n = 2,3,\dots
\label{final2}
 \end{align}
 \label{Sec:higher}

For a given slip velocity, thus a given set of amplitudes $B_n$ of the slip 
modes $V_n(\zeta)$, Eqs. \eqref{eq:BC_norm_psi} and \eqref{final2} evaluated 
for $n \geq 1$ provide a system of coupled linear equations for the last unknown 
coefficients in the stream function, $C_{n\geq 3}$ and $D_{n \geq 2}$. 
Inspection of this system reveals that it splits into two subsystems of coupled 
equations, one involving only the coefficients of even index, $n = 2 k$, and the 
other one involving only the coefficients of odd index $n = 2k + 1$ (see 
Appendix \ref{Sec:Coef}). Furthermore, from Eq. \eqref{final2} it can be 
inferred that in the case that the slip velocity is given by a pure slip mode 
$V_{n_0}(\zeta)$, i.e., $B_n = \delta_{n,n_0}$ for $n \geq 1$, then the parity 
of $n_0$ selects one of the two subsystems. If $n_0$ is even, then all the 
coefficients $C_{k}$ and $D_{k}$ of even index $k$ vanish and the stream 
function, Eq. \eqref{Ansatz1}, involves only the functions $g_k$ of odd index 
$k$, while for odd $n_0$ all the coefficients $C_{k}$ and $D_{k}$ of odd index 
$k$ vanish and the stream function, Eq. \eqref{Ansatz1}, involves only the 
functions $g_k$ of even index $k$. Finally, we note that a pure slip mode 
$V_{n_0}(\zeta)$ selects, as discussed above, either the odd or even number 
terms in the series representation of the the stream function, Eq. 
\eqref{Ansatz1}, but not only a single squirmer mode, as it is the case for the 
spherical squirmer. Accordingly, even simple distributions of active 
slip velocities on the surface of the particle can give rise to quite complex 
hydrodynamic flows around the squirmer.

With these general results, we can proceed to the discussion of 
prolate squirmers. We will focus on the usual quantities employed to 
characterize an 
axisymmetric squirmer 
\cite{lighthill1952squirming,blake1971spherical,felderhof2016stokesian}, i.e., 
the velocity $U$ of the squirmer, the magnitude $S$ of the stresslet associated 
with the squirmer (which determines the far-field hydrodynamic flow of the 
squirmer), and the general characteristics of the hydrodynamic flow 
$\mathbf{v}(\mathbf{r})$ around the squirmer. 

By combining Eqs. 
\eqref{eq:BC_tang_psi}, \eqref{squirmmodel}, \eqref{eq:orthog}, and 
\eqref{eq:def_Vn}, and noting that the 
polynomials $P_n^1(x)$ are even (odd) functions of $x$ for $n$ odd (even), one 
arrives at 
\begin{align}
 \label{eq:U_terms_Bn}
 U(\tau_0) &= \frac{\tau_0^2}{2} \sum_{n \geq 1, n~\mathrm{odd}} B_n 
 \int\limits_{-1}^{1} dx \frac{P_1^1(x) P_n^1(x)}{\tau_0^2 - x^2}\\
 &= \sum_{n \geq 1, n~\mathrm{odd}} B_n U_n(\tau_0)\,.\nonumber
\end{align}
\begin{figure}[!t]
\centering
 \includegraphics[width=0.5\textwidth]{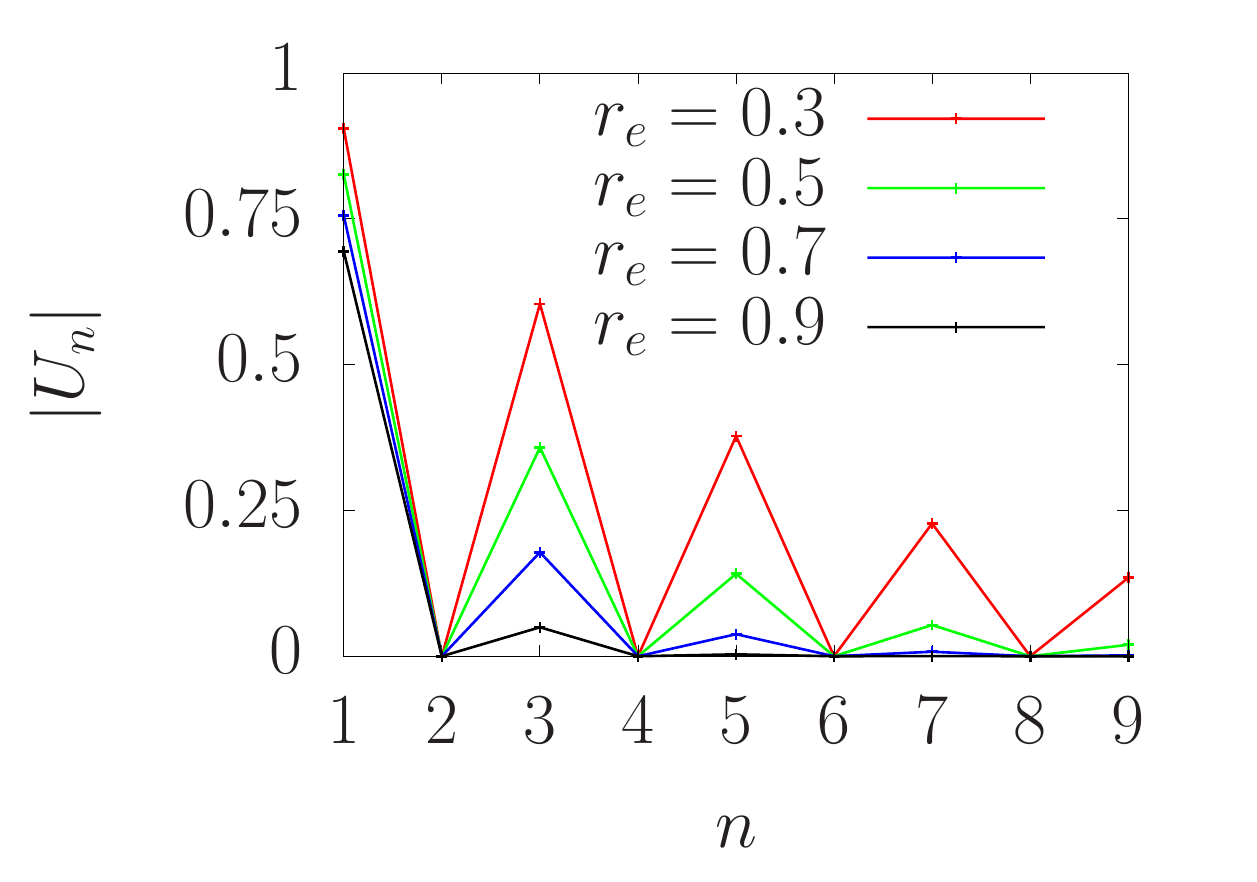}
\caption{The absolute value $|U_n|$ (Eq. \eqref{eq:U_terms_Bn}) 
of the contribution of the slip modes $n = 1,3,5,7$ to the velocity of a 
prolate squirmer for aspect ratios $r_e=0.3,0.5,0.7,0.9$. The lines 
represent only a guide to the eye.}
\label{moden}
\end{figure}
Accordingly, it follows that (a) a squirmer may exhibit self-motility (i.e., 
$U \neq 0$) only if the slip velocity involves at least an odd index $n$ 
slip mode $V_n$; (b) in contrast to the case of a spherical squirmer, for which 
the velocity is determined solely by the slip mode $n = 1$ irrespective of the 
details of the slip velocity $\mathbf{v}_s$, for a spheroidal squirmer 
all the slip modes of odd index contribute to the velocity (see also Fig. 
\ref{moden}); consequently, (c) spheroidal squirmers with $B_1 = 0$ can be 
self-motile (due to contributions from other odd index slip modes), and 
spheroidal squirmers with $B_1 \neq 0$ can yet be non-motile if the 
contributions from other slip modes of odd index 
$n$ precisely balance the contribution of the mode $B_1$ (which 
clearly pinpoints the shortcomings of a model with only two slip modes as in 
Ref. \cite{theers2016modeling}).

In terms of the dependence on the slenderness parameter $r_e$ (which determines 
the value of $\tau_0$, see Eq. \eqref{eq:tau_part}), there are two findings. 
(d) for every slip mode $n$, the contribution $|U_n|$ (in absolute 
value) is a decreasing function of $r_e$ (see Fig. \ref{moden}); second, 
(e) while at low values of the aspect ratio the contributions $|U_n|$ of the 
$n > 1$ slip modes are significant, the contributions from the modes $n \geq 3$ 
decay steeply with increasing $r_e$ and become negligible, compared to 
$|U_1|$ (which remains non-zero), as the aspect ratio $r_e$ of the spheroid 
approaches that of a sphere ($r_e \to 1^-$). This ensures a smooth transition into 
the spherical case, where, as previously mentioned, higher mode squirmers are 
not motile and $U \propto B_1$. Finally, we note that, by comparison 
with the expression in Eq. \eqref{eq:velo_bc}, one infers that the 
series in the last line of Eq. \eqref{eq:U_terms_Bn} is proportional to 
the coefficient $F_2$ in the expansion of the stream function.

In what concerns the stresslet $S(\tau)$, which (similarly to the case of 
a spherical squirmer) allows classification into pullers (positive stresslet, 
$S>0$), pushers (negative stresslet, $S<0$), and neutral swimmers (vanishing 
stresslet, $S=0$), we note that it can also be expressed in terms of the slip 
velocity $u_s(\zeta)$ as \cite{lauga2016stresslets}:
\begin{align}
 \label{eq:stresslet}
 S &=-\frac{2A\mu}{F(r_e^{-1})J(r_e^{-1})}
 \int^1_{-1}u_s(\zeta)\zeta\sqrt{\frac{r_e^{-2}(1-\zeta^2)}{\zeta^2+r_e^{-2}
(1-\zeta^2)}}d\zeta\\
 &= \sum_{n \geq 1, n~\mathrm{even}} \mu B_n S_n(\tau_0)\nonumber
\end{align}
where $A$ denotes the surface area of the spheroid and 
$$
F(x)=\frac{1}{(x^2-1)^2}\left(-3x^2+\frac{x(1+2x^2)}{\sqrt{1-x^2}}\cos^{-1}(x)\right),
$$ 
$$
J(x)=1+\frac{x^2}{\sqrt{x^2-1}}\cos^{-1}\left(\frac{1}{x}\right).
$$
\begin{figure}[!b]
\centering
 \includegraphics[width=0.5\textwidth]{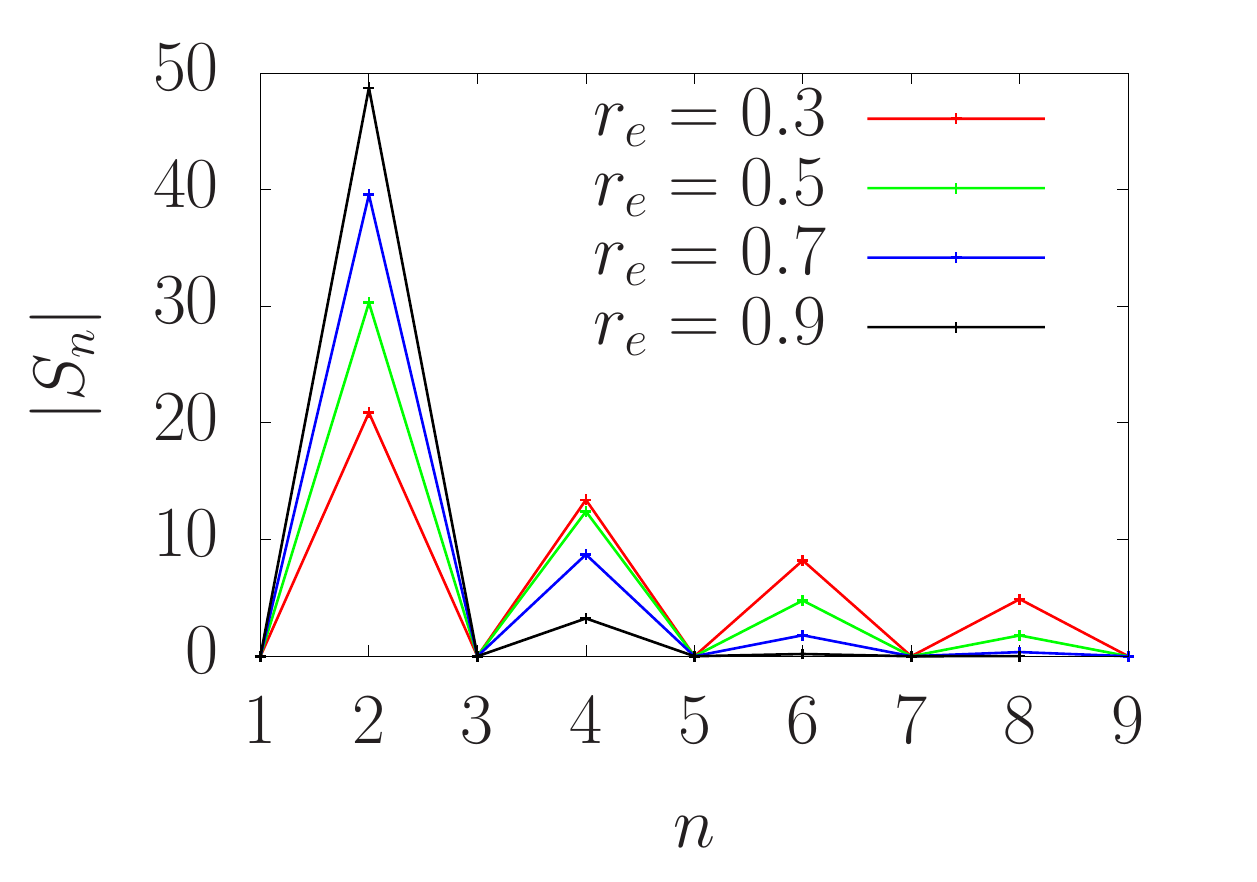}
\caption{The absolute value $|S_n|$ (Eq. \eqref{eq:stresslet}) 
of the contribution of the slip modes $n = 1,3,5,7$ to the stresslet of a 
prolate squirmer for aspect ratios $r_e=0.3,0.5,0.7,0.9$. The lines 
represent only a guide to the eye.}
\label{modenstresslet}
\end{figure}
As in the case of the velocity, there are certain significant differences 
from the case of a spherical squirmer (see Fig. \ref{modenstresslet}): (a) a 
necessary condition for a prolate squirmer to exhibit a nonvanishing stresslet 
is that at least one even ($k=2n$) mode $V_k$ contributes to the slip 
velocity; (b) as for the velocity $U$ (where more than a single mode 
contributes), the stresslet depends on all even 
squirmer modes; hence, (c) the stresslet contribution of $B_2 \neq 0$ can be 
offset or even inverted by other even, active modes $B_k \neq 0$; (d) At 
a given aspect ratio $r_e$, the contribution (in absolute value) $|S_{2n}|$ 
is a decreasing function of $n$; and (e) while for elongated shapes 
(small values $r_e$) the contributions $|S_n|$ from the slip modes $n \geq 4$ 
are significant, these contributions are steeply decreasing towards zero with 
increasing $r_e$ towards the value $1^{-}$. In contrast, $|S_2|$ is increasing 
with $r_e$. As in the case of the velocity, this behavior ensures the smooth 
transition to the case of a spherical shape, where $S \propto B_2$.
 

\begin{figure*}[!htb]
\vskip -2 \baselineskip

\hspace{-2cm}\includegraphics[width=1.1\textwidth]{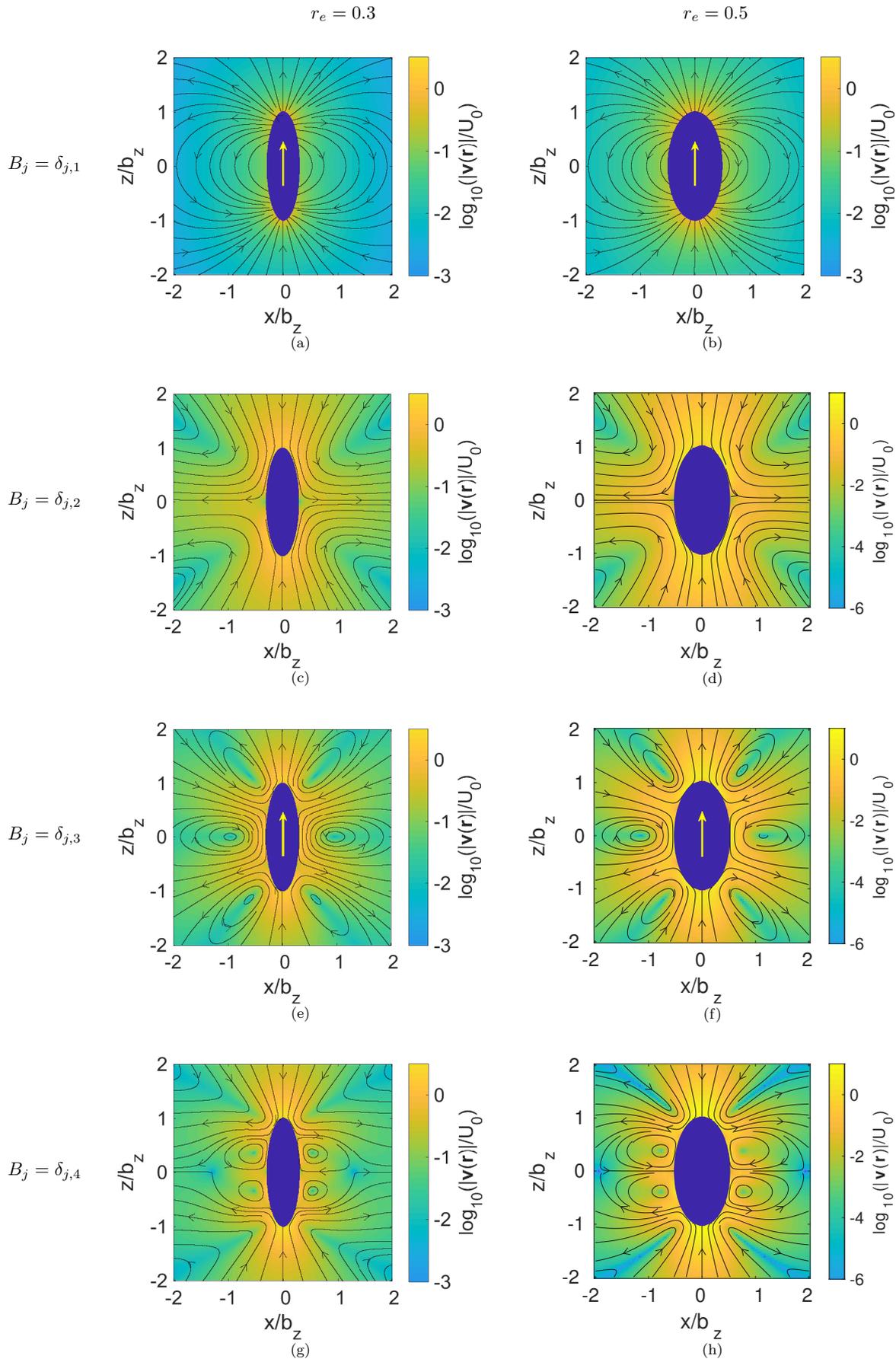}

\caption{
The flow field (streamlines and velocity magnitude (color coded 
background)) induced by a prolate squirmer with $B_n = \delta_{n,n_0}$ for (top 
to bottom) $n_0 = 1,2,3,4$ and aspect ratio $r_e=0.3$ (left column) and 
$r_e=0.5$ (right column), respectively. The results are shown in the 
laboratory frame and are obtained by using the series representation of the 
stream function (see the main text). The arrows on the particles indicate the 
directions of their motion.
}
\label{squirmers}
\end{figure*}	

Since the stresslet is, by definition, the 
amplitude of the $r^{-2}$ far-field term in the flow field (in the lab frame) 
of the squirmer, the series in the last line of Eq. \eqref{eq:stresslet} can be 
connected with one of the coefficients in the expansion of the stream function 
as follows. In the laboratory frame, which is related to the one 
($\psi_\text{particle}$) in the co-moving frame via 
\begin{equation}
\psi_\text{lab}=\psi_\text{particle}+\frac{1}{2}
U_0(1-r_e^2)(\tau^2-1)(1-\zeta^2)\,,
 \label{labframe}
\end{equation}
the slowest decaying term with the distance $\tau$ from the squirmer is 
$-\frac{C_3}{90}\text{G}_0(\tau)\text{G}_3(\zeta)$; by Eq. 
\eqref{velos}, this term leads to a contribution $\sim C_3/\tau^2$ 
to the flow. Accordingly, $S 
\propto C_3$ and the pusher or puller squirmers ($S \neq 0$) indeed exhibit 
the expected far-field hydrodynamics, while for the neutral squirmers ($S = 0$) 
the far-field flow necessarily decays at least as $\sim 1/\tau^3$.

\begin{figure}[h]
\centering
 \includegraphics[width=0.5\textwidth]{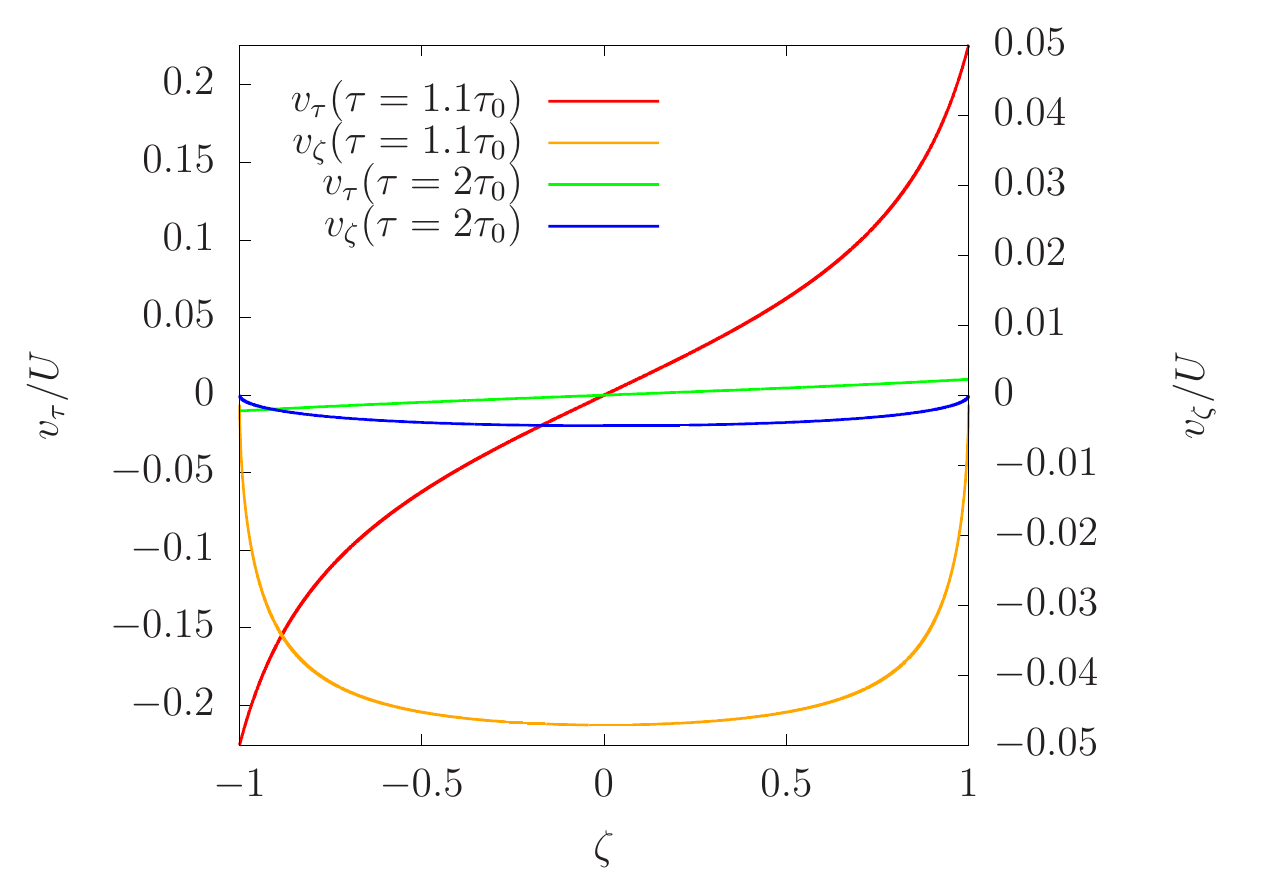}
\caption{Velocity components $v_\tau$ and $v_\zeta$ on the iso-surfaces 
$\tau=1.1\tau_0$ and $\tau=2\tau_0$ near a prolate squirmer with $B_n = 
\delta_{n,1}$.}
\label{Fig:slice1}
\end{figure}

\begin{figure}[h]
\centering
 \includegraphics[width=0.5\textwidth]{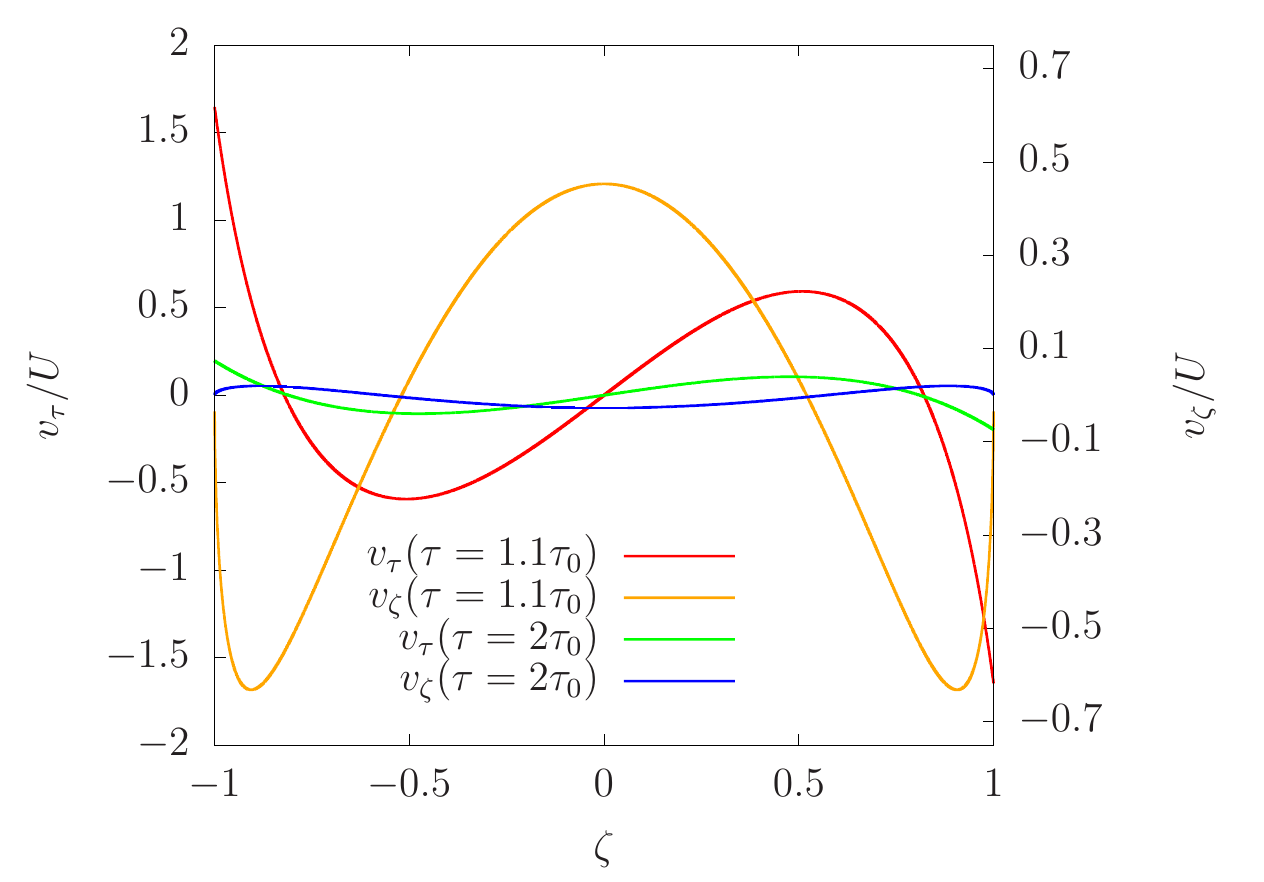}
\caption{Velocity components $v_\tau$ and $v_\zeta$ on the iso-surfaces 
$\tau=1.1\tau_0$ and $\tau=2\tau_0$ near a prolate squirmer with $B_n = 
\delta_{n,3}$.}
\label{Fig:slice2}
\end{figure}

Turning now to the flow field around the prolate squirmer, we will 
discuss separately the flow fields generated by the first few pure slip modes, 
i.e., the cases $B_n = \delta_{n,n_0}$ with $n_0 = 1, 2, 3,4$; these 
flows are shown, in the laboratory frame, in Fig. \ref{squirmers}. From the 
discussion above, we know that only a subset (either the odd index ones, if 
$n_0$ is even, or vice versa) of the terms in the series representation, Eq. 
\eqref{Ansatz1}, of the stream functions contributes to the flow. Since the 
metric factors are even functions of $\zeta$, while $G_n(\zeta)$ is an odd 
(even) function of $\zeta$ when $n$ is odd (even), the flow has the following 
fore-aft symmetries. For $n_0$ odd, the stream function involves the functions 
$G_k$ of index $k$ an 
even number, and thus $\psi(\tau,\zeta) = \psi(\tau,-\zeta)$; this implies 
$v_\tau(\tau,\zeta) = - v_\tau(\tau,-\zeta)$ and $v_\zeta(\tau,\zeta) = 
v_\zeta(\tau,-\zeta)$ (see figures \ref{Fig:slice1} and 
\ref{Fig:slice2}); i.e., the $z-$ ($x-$) flow components are fore-aft 
(anti)symmetric (see Fig. \ref{squirmers}), and, accordingly, it contributes to 
the motility because it provides a ``fore to back'' streaming. Vice 
versa, for $n_0$ an even number one has $v_\tau(\tau,\zeta) = 
v_\tau(\tau,-\zeta)$ and $v_\zeta(\tau,\zeta) = - v_\zeta(\tau,-\zeta)$, i.e., 
the $x-$ ($z-$) flow component are fore-aft symmetric (see Fig. 
\ref{squirmers}); consequently, they cannot be associated to a motile 
particle.

\begin{figure*}[!htb]
\vskip -2 \baselineskip

\hspace{-2cm}\includegraphics[width=1.1\textwidth]{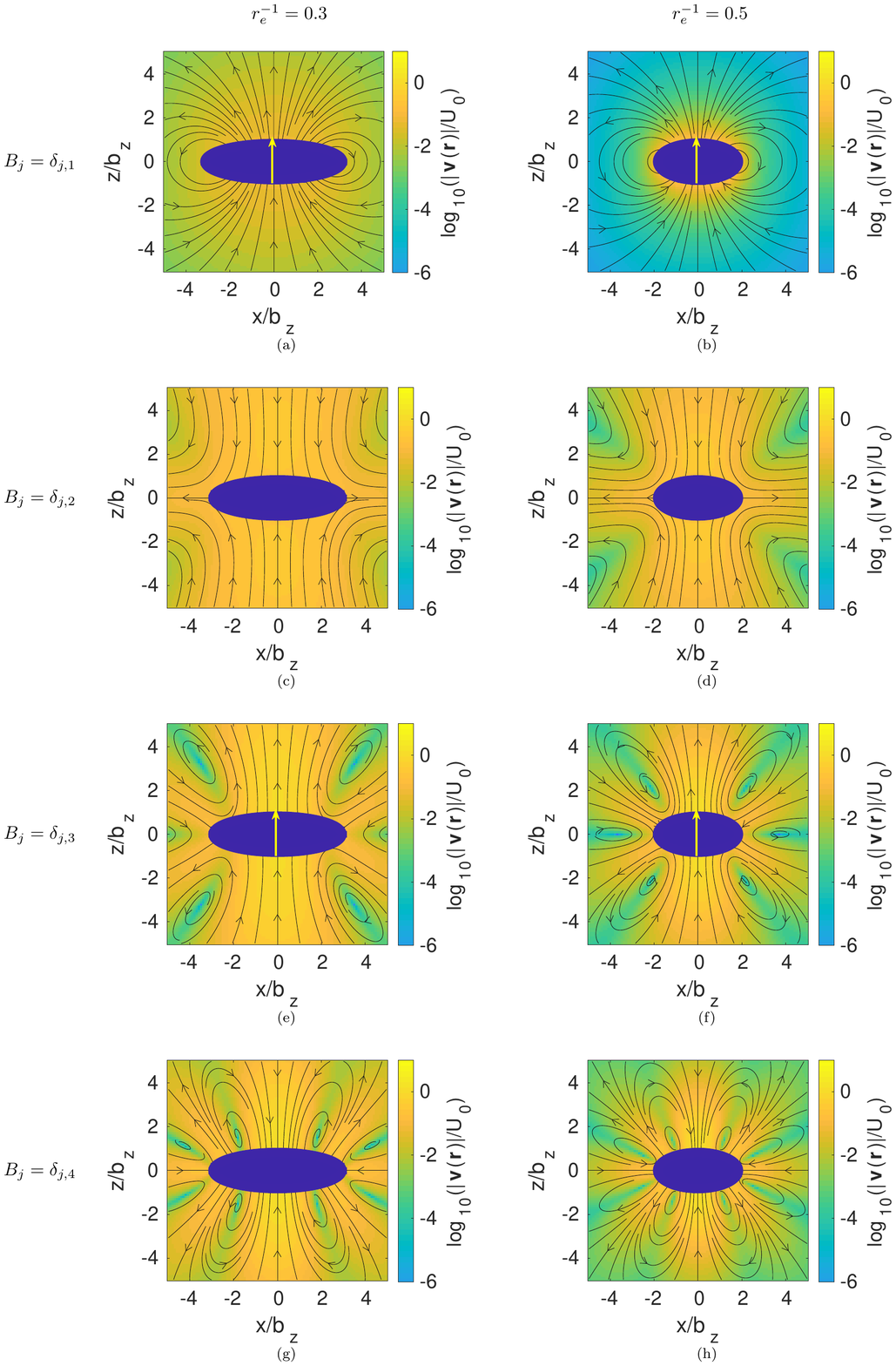}

\caption{
The flow field (streamlines and velocity magnitude (color coded 
background)) induced by an oblate squirmer with $B_n = \delta_{n,n_0}$ for 
(top to bottom) $n_0 = 1,2,3,4$ and aspect ratio $r_e^{-1}=0.3$ (left column) 
and $r_e^{-1}=0.5$ (right column), respectively. The results are shown in the 
laboratory frame and are obtained by using the mapping discussed in the main 
text and the flows of the corresponding prolate squirmers shown in Fig. 
\ref{squirmers}. The arrows on the particles indicate the directions of their 
motion.
}
\label{squirmers_obl}
\end{figure*}

\begin{figure}[h]
\centering
 \includegraphics[width=0.5\textwidth]{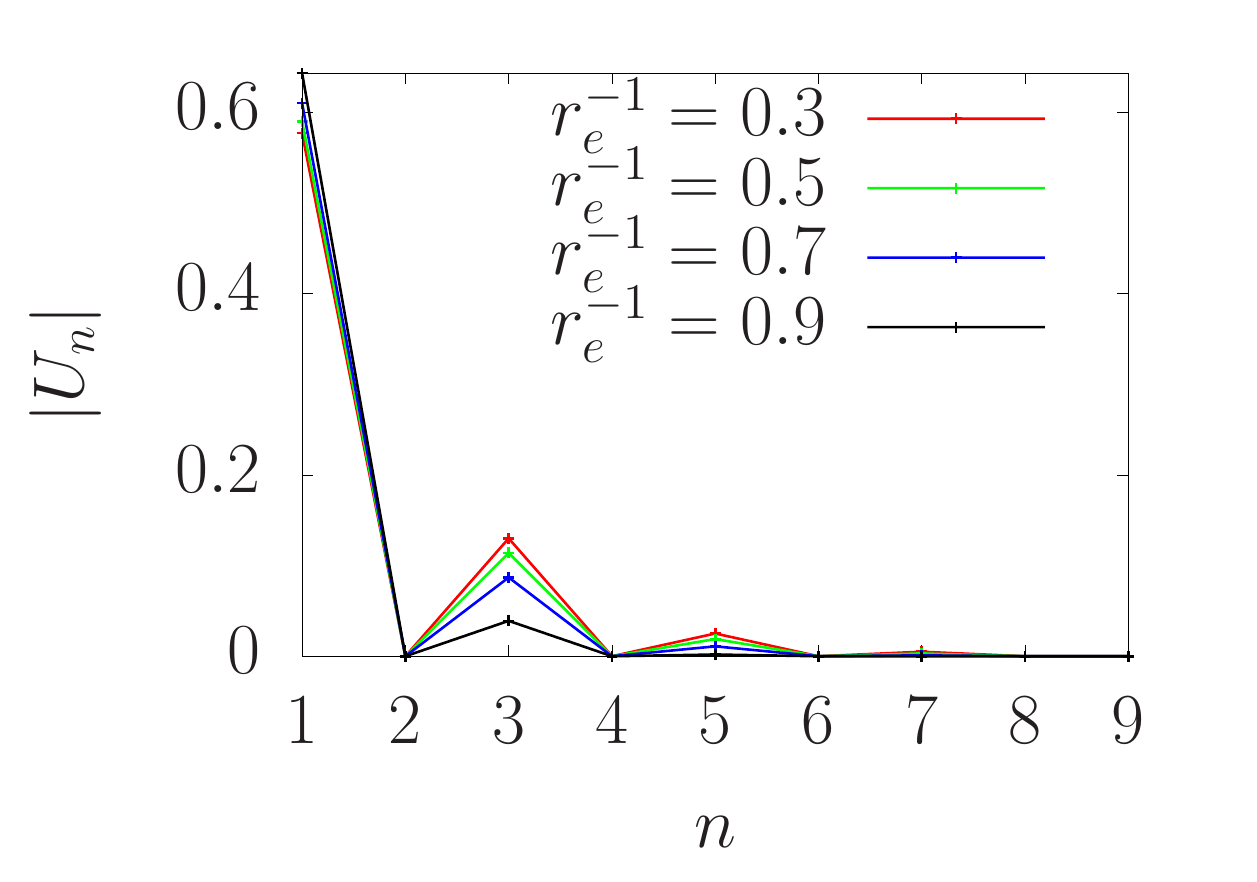}
\caption{The absolute value $|U_n|$ of the contribution of the slip modes 
$n = 1,3,5,7$ to the velocity of an oblate squirmer for aspect ratios 
$r_e^{-1}=0.3,0.5,0.7,0.9$. The lines represent only a guide to the eye.}
\label{Fig:velo_obl}
\end{figure}

\begin{figure}[h]
\centering
 \includegraphics[width=0.5\textwidth]{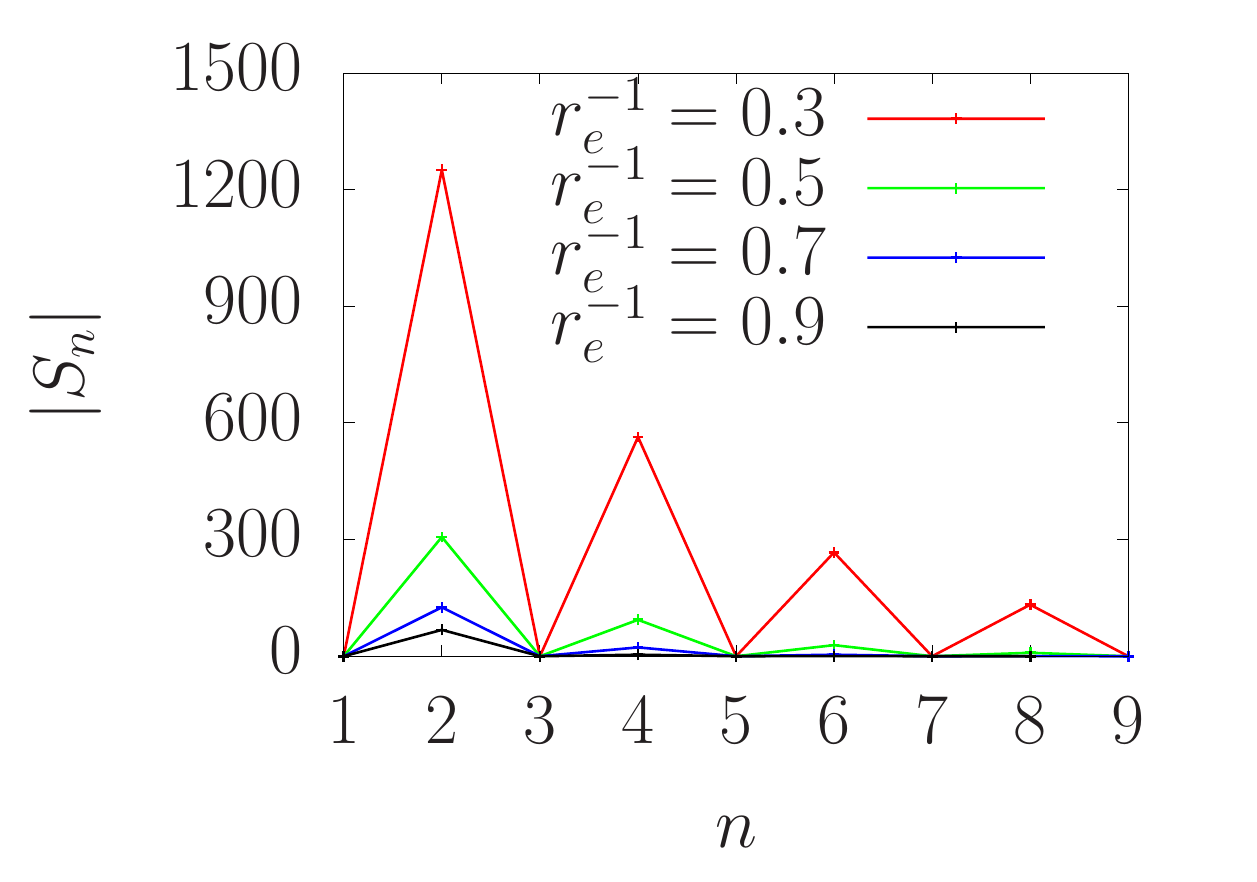}
\caption{The absolute value $|S_n|$ of the contribution of the slip modes 
$n = 1,3,5,7$ to the stresslet of an oblate squirmer for aspect ratios 
$r_e^{-1}=0.3,0.5,0.7,0.9$. The lines represent only a guide to the eye.
}
\label{Fig:stress_obl}
\end{figure}

Finally, we note that the similar analysis and results for the case of an oblate 
squirmer can be obtained from the available ones for a prolate squirmer via a 
simple transformation of the coordinates system and of the stream function 
(i.e., a mapping), as discussed in the Appendix \ref{Sec:Oblate}. As an 
illustration of using this mapping, the results shown in Fig. 
\ref{squirmers}, corresponding to a prolate squirmer, have been used to 
determine the flows around the corresponding (i.e., of 
slenderness parameters $r'_e = 1/r_e$) oblate squirmers induced by the pure 
slip modes $B_n =\delta_{n,n_0}$ with $n_0 = 1, 2, 3,4$; these are shown in 
Fig. \ref{squirmers_obl}. The analysis of the velocity $U$ and stresslet 
$S$ (see Figs. \ref{Fig:velo_obl} and \ref{Fig:stress_obl}) for oblate 
spheroids leads to conclusions that are similar with 
those drawn in the case of prolate shapes. The only significant difference is 
that now for oblate squirmers the contributions of from the higher orders 
$n$ slip modes decay to zero with \textit{decreasing} aspect ratio $r_e 
\to 1^+$; again, this ensures a smooth transition to the case of a spherical 
shape, where only $B_1$ or $B_2$ are relevant.


\newpage
\begin{figure*}[h]
\vskip -4 \baselineskip

\hspace{-2cm}\includegraphics[width=1.1\textwidth]{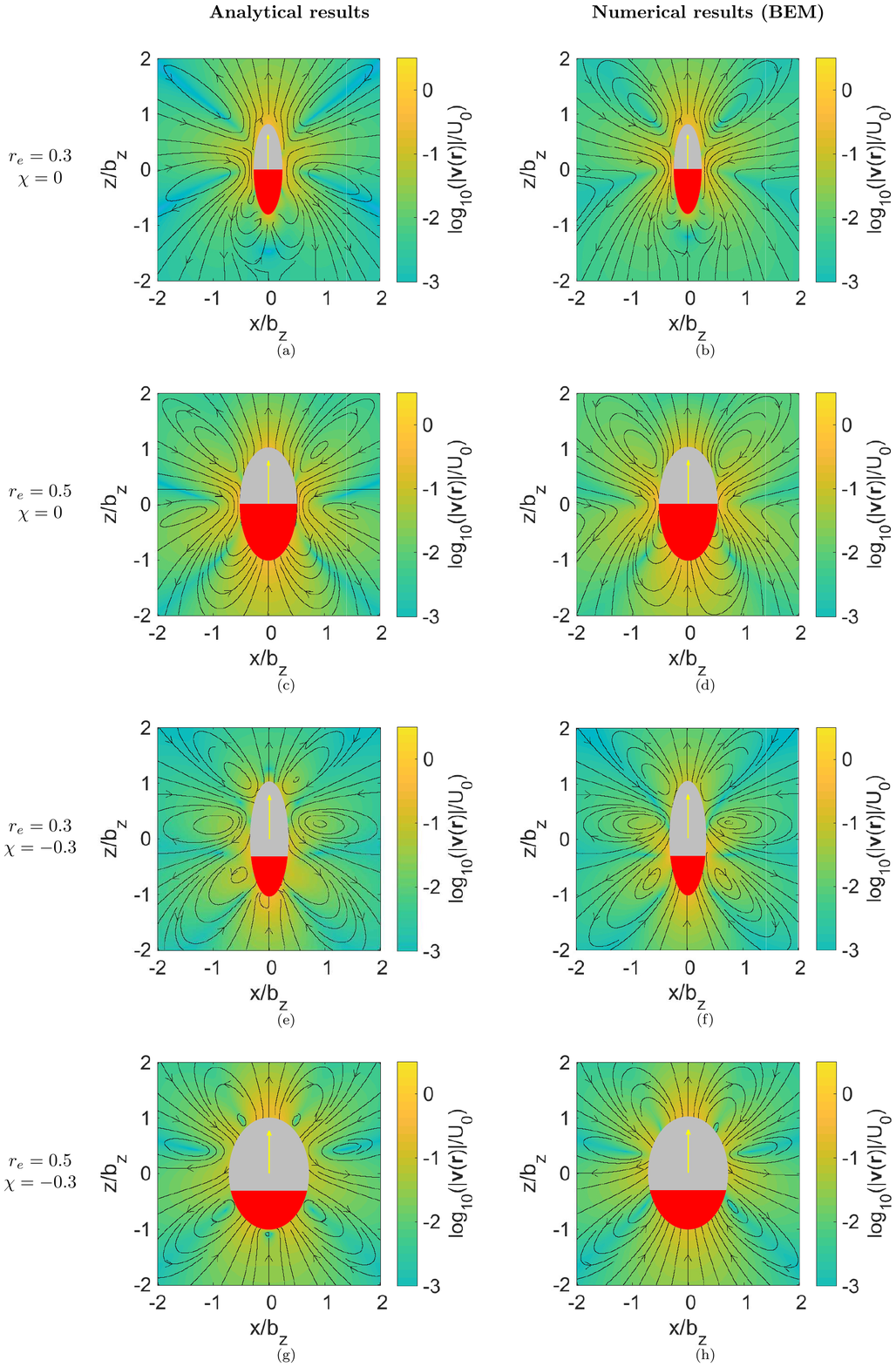}

\caption{
The flow field (streamlines and velocity magnitude (color coded 
background)) induced by a chemically active prolate particle moving by 
self-phoresis, calculated either analytically via the stream function (left 
column) or numerically, i.e., by directly solving the corresponding 
Laplace and Stokes equations using the BEM \cite{BEM} (right). The 
results shown correspond to the cases of half ($\chi=0$, top two rows) and less 
than half ($\chi=-0.3$, bottom two rows) coverage, and two values, $r_e=0.3$ 
and $r_e=0.5$, of the aspect ratio, respectively. The arrows on the particles 
indicate the directions of their motion. The red area depicts the chemically 
active region. The motion of the particle and 
the direction of the flow corresponds to the choice $b < 0$; the 
characteristic velocity $U_0$ is defined as in Ref. \cite{popescu2010phoretic} 
}
\label{chempar1}
\end{figure*}

\subsection{Self-phoretic particle\label{Sec:Phoretic}}
Squirmers with a wide range of active slip modes occur naturally in the 
context of model self-phoretic particles. One of the often employed realizations 
of such systems consists of micrometer-sized silica or polystyrene spherical 
particles partially coated with a Pt layer and immersed in an aqueous peroxide 
solution \cite{Howse2007,Baraban2012,Simmchen2016}. The catalytic decomposition 
of the peroxide at the Pt side creates gradients in the chemical composition of 
the suspension; as in the case of classic phoresis \cite{anderson1989colloid}, 
these gradients, in conjunction with the interaction between the colloid and 
the various molecular species in solution, give rise to self-phoretic motility 
\cite{Golestanian2005}. The mechanism of steady-state motility can be 
intuitively understood in terms of the creation of a so-called phoretic slip 
velocity tangential to the surface of the particle 
\cite{golestanian2007designing,anderson1989colloid,Derjaguin1966}. For such 
chemically active, axi-symmetric, spherical particles in unbounded solutions, 
the approximation of phoretic slip velocity leads to a straightforward mapping 
\cite{Michelin2014,Popescu2018} onto a squirmer model; accordingly, the 
translational velocity of the particle and the hydrodynamic flow around the 
particle in terms of the phoretic slip can be directly inferred from the 
corresponding results in Ref. \cite{blake1971spherical}.

A similar mapping can be developed from a spheroidal, self-phoretic 
colloid to a spheroidal squirmer model for spheroidal, chemically active 
colloids. Following Ref. \cite{popescu2010phoretic} the slip velocity 
of such particles can be written as
\begin{align}
 v_\text{S}(\bold{r}_\text{P})=\sum_{l\geq0}\frac{b}
 {c} 
c_l(\chi) \text{Q}_l(\tau_0) \text{P}^1_l(\zeta)
(\tau_0^2-\zeta^2)^{-\frac{1}{2}},
 \label{vschem}
\end{align}
where $b$ is the so-called phoretic mobility (for simplicity, here assumed 
to be a constant) over the surface of the particle, and $\text{Q}_l(\tau)$ denotes 
the Legendre polynomial of the second kind \cite{Abramowitz}. The coverage $\chi = h-1$ 
is defined in terms of the height of the active cap (measured, from the 
bottom apex, in units of $b_z$), i.e., $\chi = -1$ corresponds to a 
chemically inactive spheroid and $\chi = +1$ corresponds to all the surface 
being active. The coverage dependent coefficients $c_l(\chi)$ (see Ref. 
\cite{popescu2010phoretic}) describe the decoration of the surface of 
the particle by the chemically active element (e.g., Pt in the example 
discussed above) as an expansion in Legendre polynomials $\text{P}_l(\zeta)$. 
Knowledge of these parameters allows us to formulate a slip velocity boundary 
condition similar to \eqref{final2}, i.e.
 \begin{align}
 \frac{\partial g_{n}(\tau)}{\partial\tau}\Big|_{\tau=\tau_0} 
= \tau_0c^2n(n-1)\tilde{B}_{n-1}\,;
\label{final3}
 \end{align}
by identifying the effective squirmer modes $\tilde{B}_l 
=\frac{b}{c\cdot\tau_0}c_l\text{Q}_l(\tau_0)$, the desired mapping is 
achieved. As an illustration of this mapping, we show in Fig. 
\ref{chempar1} (left column) the flow fields for particles with parameters 
$r_e=0.3,0.5$ and $\chi=0,-0.3$, respectively, and compare with the 
corresponding results obtained by direct numerical solutions obtained using BEM 
\cite{BEM}. (Note that, if necessary, the accuracy of the analytical 
estimate can be systematically improved simply by increasing the order of the 
truncation in the series expansion of the stream function (see the 
Appendix \ref{Sec:Conv})).

\section{Summary and conclusion\label{Sec:Summary}}

We have studied in detail the most general axi-symmetric spheroidal squirmer 
and we have shown that, in analogy with the situation for spherical shapes 
\cite{Michelin2014}, model chemically active, self-phoretic colloids can be 
mapped onto squirmers. By using the semiseparable ansatz derived in Ref. 
\cite{dassios1994generalized} for the stream function, and representing the 
active slip (squirming) velocity of the squirmer in a suitable basis (Eq. 
(\ref{squirmmodel}), chosen such that in the limit of a spherical shape it 
smoothly transforms into the usually employed expansion for the classical 
spherical squirmer \cite{blake1971spherical}, the velocity of the squirmer, 
the stresslet of the squirmer, and the hydrodynamic flow around the squirmer 
have been determined analytically (Sec. \ref{Sec:Squirmer}). The 
corresponding series representations have been validated by cross-checking 
against direct numerical calculations, obtained by using the BEM, of the flow 
around the squirmer (Appendix \ref{Sec:Num1}).

The main conclusion emerging from the study is that for spheroidal squirmers 
(or self-phoretic particles) the squirming modes beyond the second are, in 
general, as important as the first two ones in what concerns the contributions 
to the velocity and stresslet of the particle (and, implicitly, to the flow field 
around the particle, even in the far-field). The velocity is contributed by all 
the odd-index components (but none of the even-index ones) of the slip 
velocity; accordingly, in contrast with the case of spherical squirmers, it is 
possible to have spheroidal squirmers with a non-zero first mode and yet not 
motile, as well as ones missing the first slip mode and yet motile. Similarly, 
the stresslet value is contributed by all the even-index slip modes; thus, 
distinctly from the case of spherical squirmers, one can have pushers/pullers 
even if the second slip mode is vanishing, as well as neutral squirmers in 
spite of a non-vanishing second slip mode. Finally, even a single slip mode 
leads to a large number of non-vanishing terms in the series expansion of the 
stream function, and thus spheroidal squirmers with simple distributions of 
slip on their surface can lead to very complex flows around them (see Figs. 
\ref{squirmers} and \ref{squirmers_obl}).

This raises the interesting speculative question as whether the spheroidal 
shape is providing an evolutionary advantage; i.e., with small 
modifications of the squirming pattern, e.g., switching from a sole $B_1$ 
mode to a sole $B_3$ mode, a microrganism could maintain its velocity 
unchanged but dramatically alter the topology of the flow around it 
(compare the first and third rows in Fig. 4). In other words, compared to a 
squirmer with spherical shape, for which multiple modes must be 
simultaneously activated in order to change the structure of the flow, a 
spheroidal squirmer possesses simple means for acting in hydrodynamic 
disguise, which can be advantageous as either predator or prey.

\begin{appendices}

\subsection{Oblate spheroids \label{Sec:Oblate}}
The similarity between oblate and prolate spheroids allows us to obtain the 
flow field around an oblate microswimmer of aspect ratio $r_e' > 1$ as a 
series in the oblate coordinates by using a mapping from the results, in 
prolate coordinates, corresponding to a prolate microswimmer with aspect ratio 
$r_e = 1/r_e' < 1$ (and vice versa).

The oblate spheroidal coordinate system is defined by

\begin{align}
\lambda = \frac{1}{2\bar{c}}(\sqrt{x^2+y^2+(z-i\bar{c})^2}+ 
&\sqrt{x^2+y^2+(z+i\bar{c})^2}\large{)}\,,\nonumber\\
\zeta= \frac{1}{-2i\bar{c}}(\sqrt{x^2+y^2+(z-i\bar{c})^2}-&\sqrt{x^2+y^2+(z+i\bar{c})^2}),\nonumber\\
\varphi = \arctan & \left(\frac{y}{x}\right)\,,\nonumber
\end{align}

with $0\leq\lambda\leq\infty$ and $\bar{c}=\sqrt{r_e'^2-1}$, and the 
corresponding Lam{\'e} metric coefficients are given by
\begin{align}
\label{eq:metric_coef_oblate}
& h_\zeta =\bar{c}\frac{\sqrt{\lambda^2+\zeta^2}}{\sqrt{1-\zeta^2}}\,, 
~~~h_\lambda=\bar{c}\frac{\sqrt{\lambda^2+\zeta^2}}{\sqrt{\lambda^2+1}}\,, \\
& h_\varphi = \bar{c} \sqrt{\lambda^2+1}\sqrt{1-\zeta^2}\,.\nonumber
\end{align}
Noting that the oblate coordinates and the metric factors 
can be obtained from the expressions of the corresponding prolate 
coordinates via the transformations \cite{dassios1994generalized}
\begin{equation}
\tau\rightarrow i\lambda \qquad c \rightarrow -i\bar{c} \,,
\end{equation}
one concludes that the equation and boundary conditions obeyed by the 
stream 
function in oblate coordinates in the domain outside the oblate of aspect ratio 
$r_e' = 1/r_e$ can be obtained, by using the same transformation, from the ones 
in prolate coordinates outside a prolate of aspect ratio $r_e$. Accordingly, it 
follows that $\psi_\text{obl,$r_e'=1/r_e$}(\lambda,\zeta) = 
\psi_\text{pro,$r_e$}(\tau = i\lambda,\zeta)$.
The flow field then follows as the curl of the stream function, i.e.,
\begin{equation}
\bold{v}_\text{obl,$r_e'$}(\lambda,\xi,\varphi)= \nabla \times 
\left(\frac{ \psi_\text{obl,$r_e'$} 
(\lambda,\xi)}{h_\varphi} \,\,\bold{e}_\varphi \right) .
\end{equation}

\subsection{Gegenbauer functions\label{Sec:Ortho}}
The Gegenbauer functions of first kind $G_n$, are also known as the 
Gegenbauer polynomials $C^\alpha _n$ with parameter $\alpha = -1/2$ \cite{Abramowitz}; 
they are defined in terms of the Legendre polynomials $P_n$ as
\begin{subequations}
 \begin{equation}
 G_n(x)=\frac{1}{2n-1}\left(P_{n-2}(x)-P_n(x)\right)\;,
\end{equation}
\end{subequations}
and fulfill the orthogonality relation
\begin{equation}
\int_{-1}^1\frac{G_n(x)G_m(x)}{1-x^2}dx=\frac{2}{n(n-1)(2n-1)}\delta_{n,m}\quad 
n,m\ge2\,.
\end{equation}
The Gegenbauer functions can be related to the associated Legendre 
polynomials $\text{P}^1_n$,
\label{Sec:Ident}
which are defined in terms of the Legendre polynomials $P_n$ as
\begin{align}
(1-x^2)^{1/2}\text{P}^1_l = -(1-x^2)\frac{\text{d}}{\text{d}x}P_l(x).\nonumber
\end{align}
Using the relations 
$\frac{x^2-1}{n}\frac{\text{d}}{\text{d}x}P_n(x)=x\text{P}_n(x)-\text{P }_{n-1} 
(x)$ and $(n+1)\text{P}_{n+1}(x)=(2n+1)x\text{P}_n(x)-n\text{P}_{n-1}(x)$, 
one arrives at the following relation between the $l$-th 
associated Legendre polynomial and the $l+1$-th Gegenbauer function.
\begin{align}
(1-x^2)^{1/2}\text{P}^1_l =-(l^2+l)\text{G}_{l+1}(x)\nonumber
\end{align}

\subsection{Decoupling of the even- and odd-index modes in the stream 
function expansion \label{Sec:Coef}}

The coefficients $C_n$ and $D_n$ appear in the functions $g_k(\tau)$, 
entering the series expansion of the stream function at various indexes $k$ 
(e.g., $C_{n \geq 4}$ appears at both $k = n-2$ and $k = n$); thus the 
infinite system of linear equations is strongly coupled. However, since the 
even and odd terms are not entering the same equations, the system splits into 
two decoupled subsystems, which are solved by using different methods.

The first subsystem consists of the even numbered terms in the series 
expansion of $\psi(\tau,\zeta)$ and the set of conditions (Eqs. 
\eqref{eq:forc_bal_psi}, \eqref{eq:velo_bc}, \eqref{eq:BC_norm_psi} and 
\eqref{final2}). Because each of the Eqs. \eqref{eq:forc_bal_psi} and 
\eqref{eq:velo_bc} fix one of the even index coefficients (i.e., $C_2$ and 
$F_2$), the Eqs. \eqref{eq:BC_norm_psi} and \eqref{final2} evaluated at $n=2$ 
involve only two unknowns, $C_4$ and $D_2$, and thus can be solved as a 
sub-subsystem. With $C_4$ and $D_2$ known, the rest of the coefficients, up to 
the order $N_{max}$ at which the system is truncated (i.e., $B_k$ is set to 
zero for $k > N_{max}$), are solved iteratively by noting that Eqs. 
\eqref{eq:BC_norm_psi} and \eqref{final2} evaluated at $k = n$ involve 
only two unknowns, the rest of the coefficients being already 
determined in the previous iterations up to $k = n-2$.

The second subsystem includes all odd-index terms in the series expansion 
of $\psi(\tau,\zeta)$ and Eqs. \eqref{eq:BC_norm_psi} and \eqref{final2}. 
Complementing the previous case, only the even squirmer modes contribute. 
However, unlike in the case of the first subsystem, there are no lower 
level decouplings; accordingly, this subsystem is solved in the standard manner 
by truncation at the cut-off $N_{max}$ (above which all the coefficients are 
set to zero) and inversion of the resulting finite system of linear equations.

The choice of $N_{max}$ is done by varying the value of the cut-off and 
testing the changes in the coefficients. For the cases we analyzed in this 
work, a value $N_{max} = 16$ was found to be sufficient.

\begin{figure*}
\begin{tabular}{@{}c@{ }c@{ }c@{ }c@{}}
\subfigure[]{\includegraphics[width=0.48\textwidth]{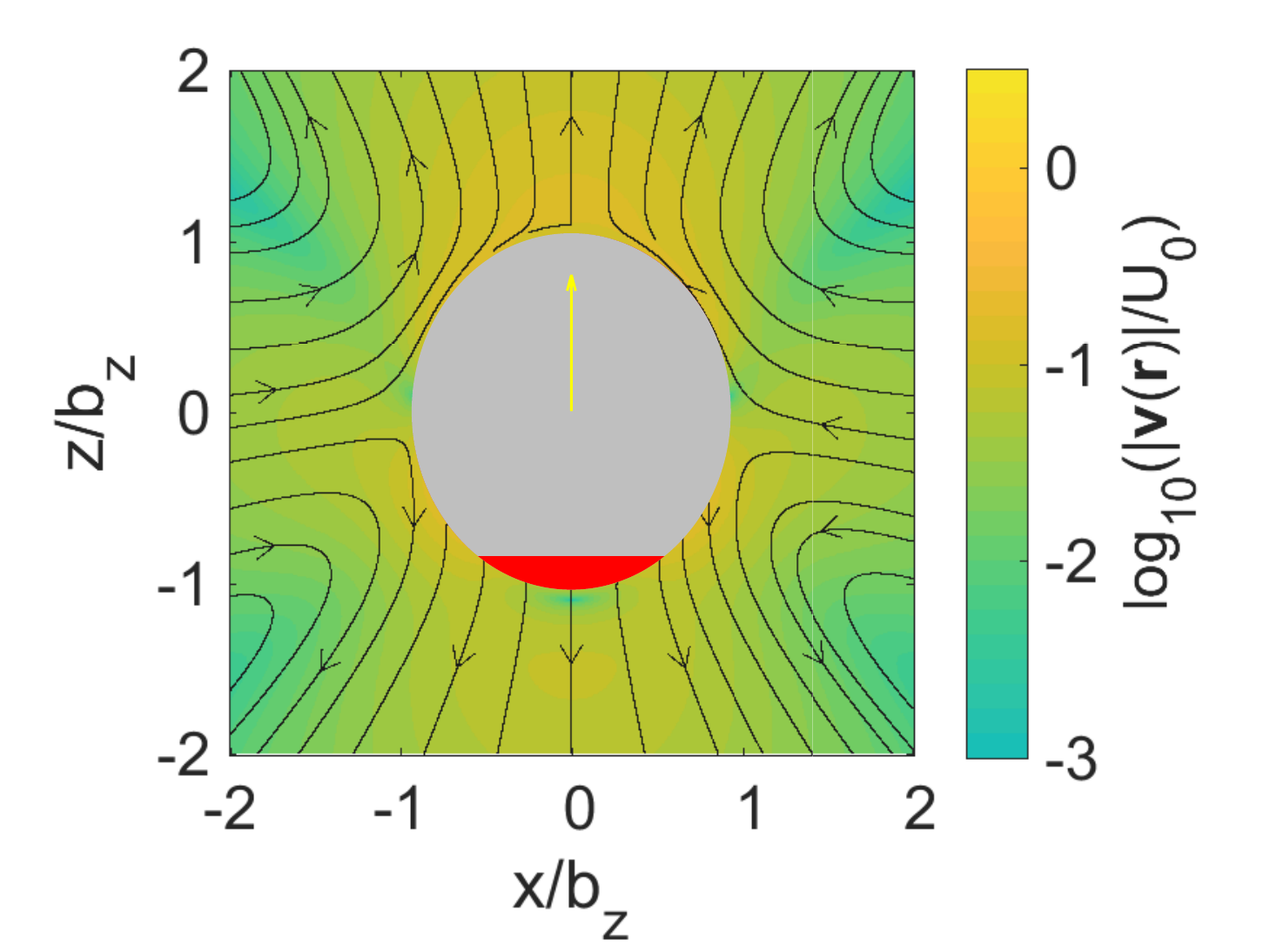}}&
\subfigure[]{\includegraphics[width=0.48\textwidth]{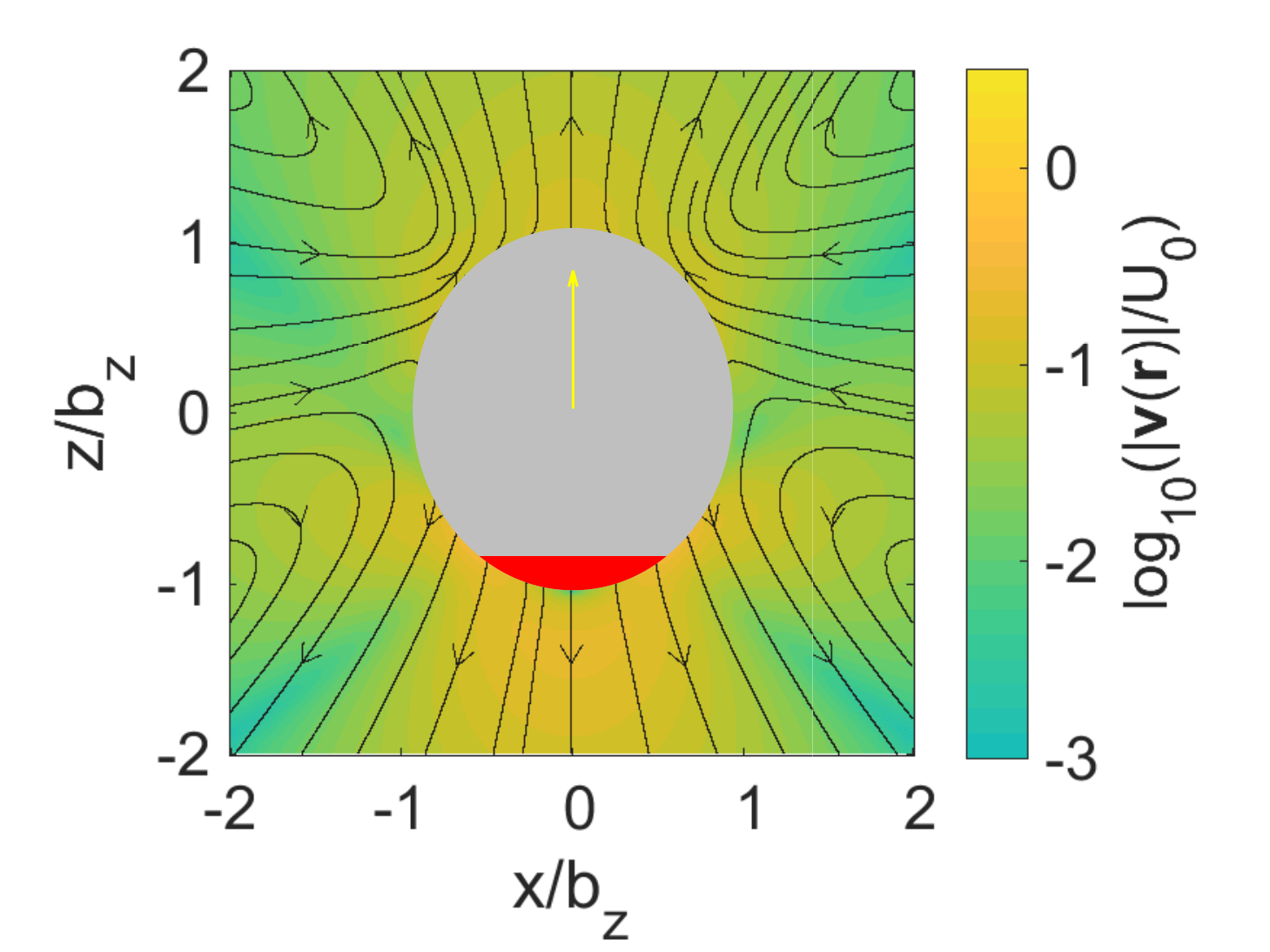}}\\[-1ex]
\end{tabular}
\caption{The flow field around a self-phoretic particle ($\chi=0.8$, 
$r_e=0.9$) when keeping the first (a) two and (b) eight terms (effective 
squirmer modes) in the series expansion of the stream function. 
The red area depicts the chemically active region.
}
\label{selfmoden}
\end{figure*}

\subsection{Example of effects of too strong truncations for the case
self-phoretic particles \label{Sec:Conv}}

To show the importance of the higher orders in the case of a self-phoretic 
swimmer, we compare the analytic results for keeping different numbers of 
squirmer modes $B_n$ in figure \ref{selfmoden}. Even for an aspect ratio 
$r_e=0.8$ (i.e., close to a spherical shape), the higher orders have significant 
influence on the flow field around the particle: e.g., compare the occurrence of 
a region of high magnitude flow near the point $(0,-1.5)$ (behind the particle) 
instead of the correct location at near the point $(0,1.5)$ (in front of 
the particle).

\subsection{Quantitative comparison to BEM\label{Sec:Num1}}
\begin{figure*}[!htb]
\vskip -4 \baselineskip

\makebox[\textwidth][c]{\includegraphics[width=1.15\textwidth]{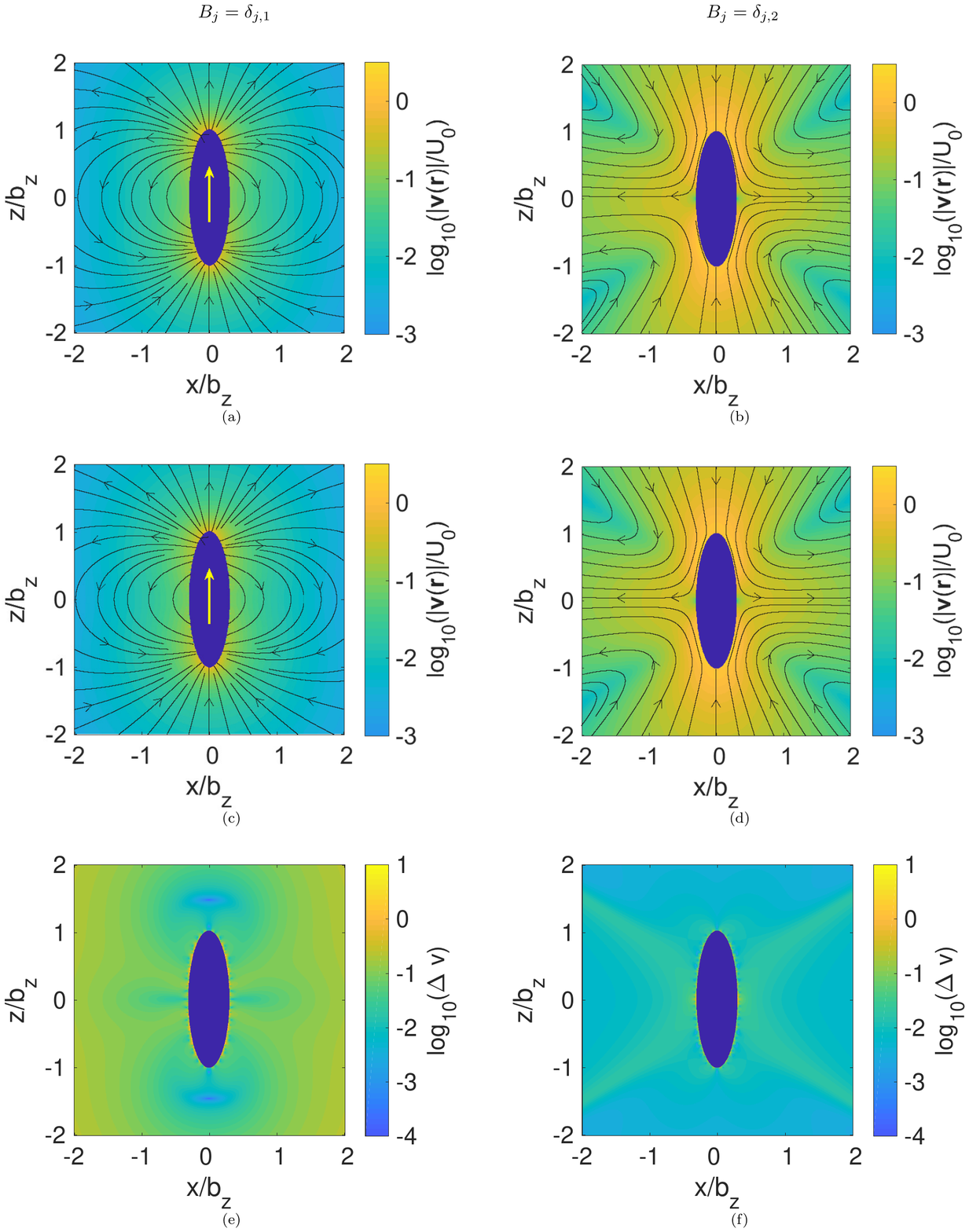}}
\vskip -4 \baselineskip

\caption{The flow field around a prolate squirmer with $r_e=0.3$; in the 
left column only the first slip mode is active, and on the right column only 
the second slip mode is active. In both cases the analytical results are shown 
in the top row, the numerical results (BEM \cite{BEM}) in the middle row, and 
the relative error between the two (Eq. \eqref{error}) in the bottom row.}
\label{B1/2}

\end{figure*}

\begin{figure*}[!htb]
\vskip -4 \baselineskip

\makebox[\textwidth][c]{\includegraphics[width=1.15\textwidth]{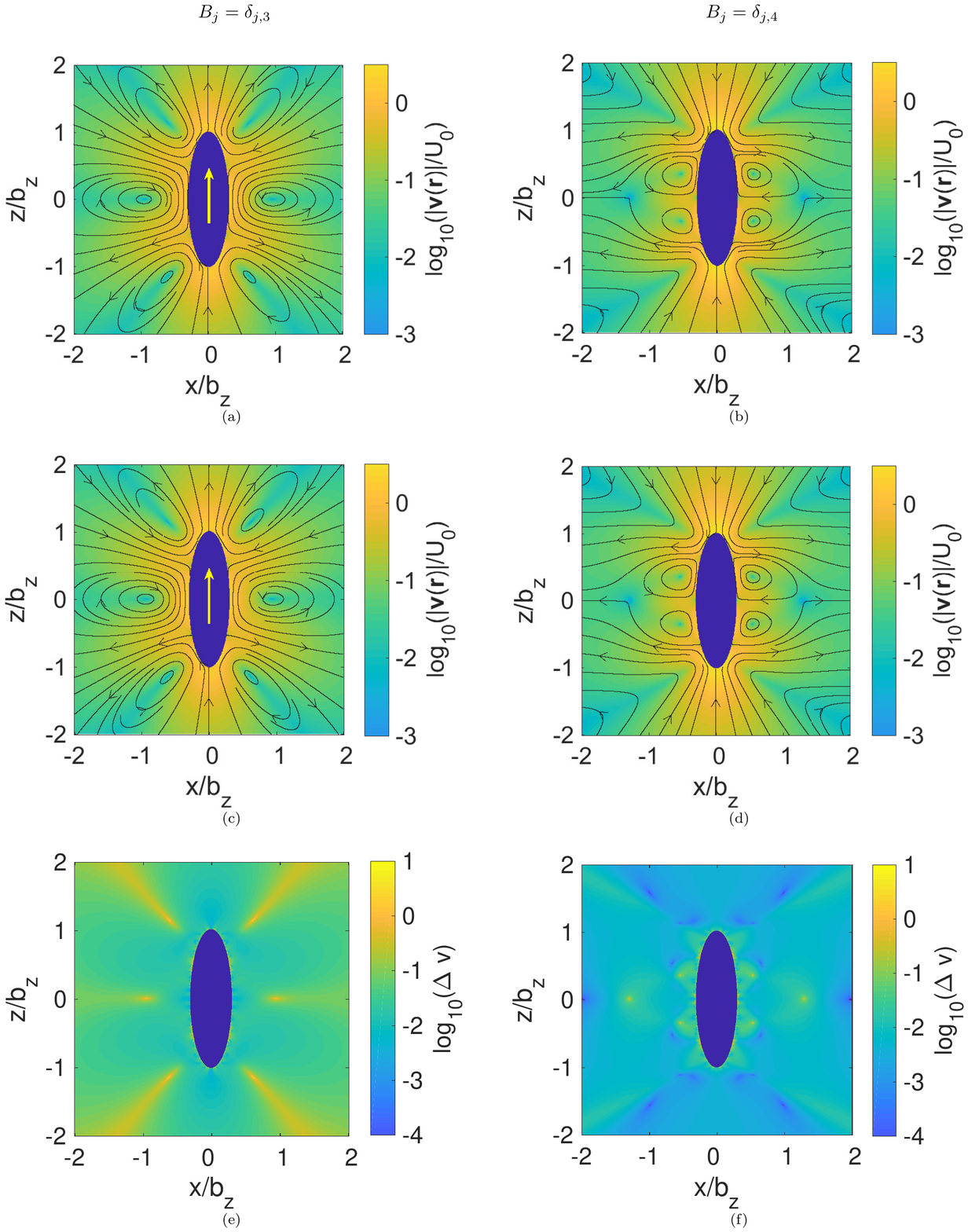}}
\vskip -4 \baselineskip
\caption{The flow field around a prolate squirmer with $r_e=0.3$; in the 
left column only the third slip mode is active, and on the right column only 
the fourth slip mode is active. In both cases the analytical results are shown 
in the top row, the numerical results (BEM \cite{BEM}) in the middle row, and 
the relative error between the two (Eq. \eqref{error}) in the bottom row.}
\label{B3/4}
\end{figure*}

\begin{figure*}[h]
\vskip -2 \baselineskip

\makebox[\textwidth][c]{\includegraphics[width=1\textwidth]{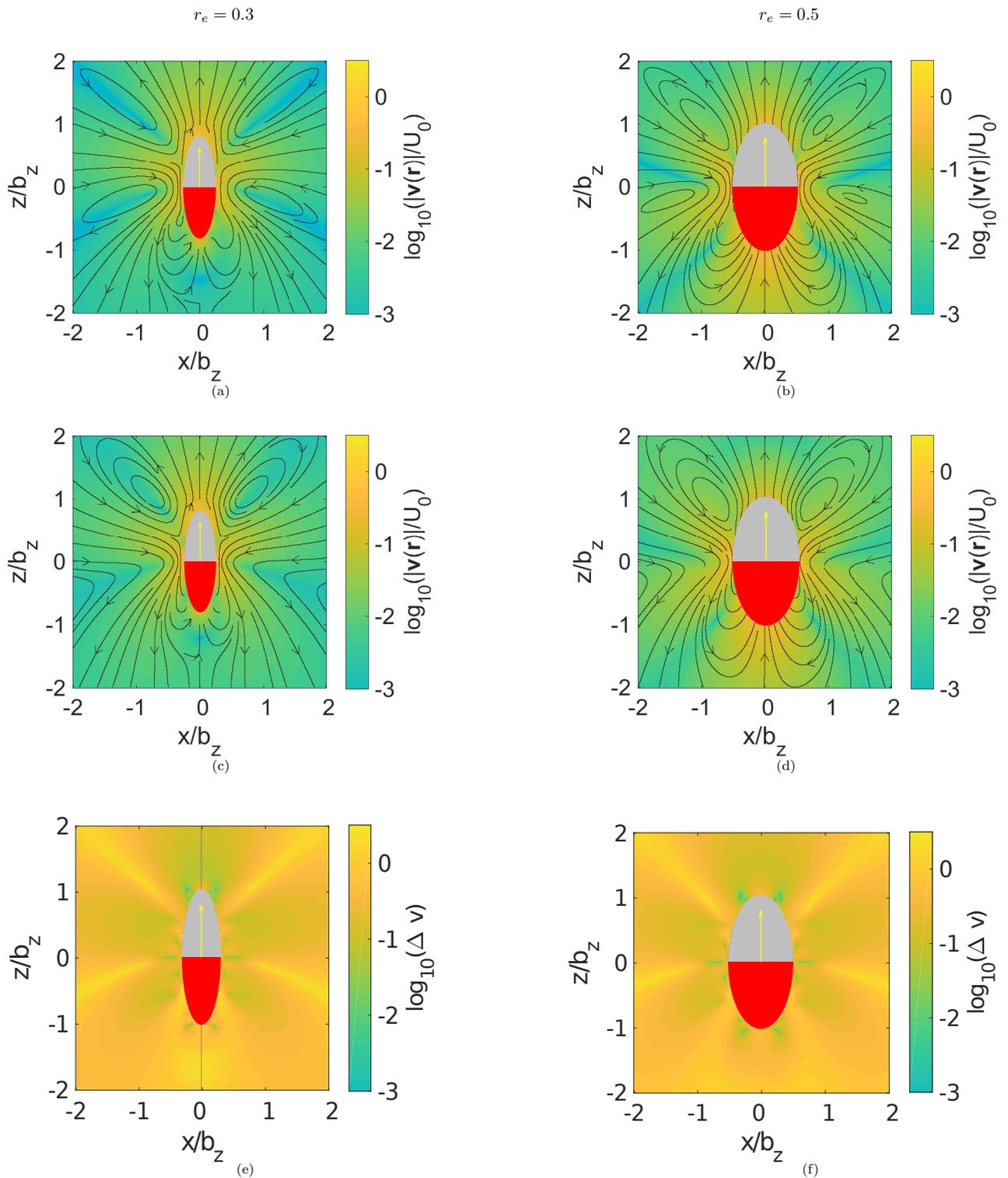}}
\vskip -4 \baselineskip
\caption{The flow field around half covered ($\chi=0$) self-phoretic 
particles with aspect ratio $r_e=0.3$ (left) and $r_e=0.5$ (right). In both 
cases the analytical results are shown 
in the top row, the numerical results (BEM \cite{BEM}) in the middle row, and 
the relative error between the two (Eq. \eqref{error}) in the bottom row. 
The red area depicts the chemically active region.
}
\label{chempar2}
\end{figure*}

In figures \ref{B1/2} - \ref{chempar2} we present a more detailed 
comparison of the results obtained by using the series representation (top row) 
with numerical results obtained by using the BEM to directly solve the 
governing equations (second row). In addition, the relative error, defined as
\begin{equation}
\Delta v= \frac{\sqrt{(v_{x,ana}-v_{x,num})^2+
(v_{z,ana}-v_{z,num})^2}}{\frac{1}{2}(\sqrt{v_{x,ana}^2+v_{z,ana}^2}+\sqrt{v_{x,
num}^2+v_{z,num}^2})}
 \label{error}
\end{equation}
is shown in the bottom row. The comparison confirms the expected quantitative 
agreement, with regions of significant relative error ($\Delta v 
> 10^{-1}$) corresponding precisely to the regions where the flow is anyway very weak 
($|\frac{\bold{v}/\bold{r})}{U}|<10^{-3}$).
\vskip -4 \baselineskip

\end{appendices}
\clearpage

\bibliographystyle{baen}
\bibliography{References}

\begin{thebibliography}{10}
\providecommand{\bibAnnoteFile}[1]{%
  \IfFileExists{#1}{\begin{quotation}\noindent\textsc{Key:} #1\\
  \textsc{Annotation:}\ \input{#1}\end{quotation}}{}}
\providecommand{\bibAnnote}[2]{%
  \begin{quotation}\noindent\textsc{Key:} #1\\
  \textsc{Annotation:}\ #2\end{quotation}}

\bibitem{pedley1992hydrodynamic}
T.J. Pedley and J.O. Kessler.
\newblock Hydrodynamic phenomena in suspensions of swimming microorganisms.
\newblock \emph{Annual Review of Fluid Mechanics} 24, 313--358 (1992).
\bibAnnoteFile{pedley1992hydrodynamic}

\bibitem{guasto2012fluid}
J.S. Guasto, R.~Rusconi, and R.~Stocker.
\newblock Fluid mechanics of planktonic microorganisms.
\newblock \emph{Annual Review of Fluid Mechanics} 44, 373--400 (2012).
\bibAnnoteFile{guasto2012fluid}

\bibitem{lighthill1952squirming}
M.J. Lighthill.
\newblock On the squirming motion of nearly spherical deformable bodies through
  liquids at very small Reynolds numbers.
\newblock \emph{Communications on Pure and Applied Mathematics} 5, 109--118
  (1952).
\bibAnnoteFile{lighthill1952squirming}

\bibitem{blake1971spherical}
J.R. Blake.
\newblock A spherical envelope approach to ciliary propulsion.
\newblock \emph{Journal of Fluid Mechanics} 46, 199--208 (1971).
\bibAnnoteFile{blake1971spherical}

\bibitem{zottl2014hydrodynamics}
A.~Z{\"o}ttl and H.~Stark.
\newblock Hydrodynamics determines collective motion and phase behavior of
  active colloids in quasi-two-dimensional confinement.
\newblock \emph{Physical Review Letters} 112, 118101 (2014).
\bibAnnoteFile{zottl2014hydrodynamics}

\bibitem{gotze2010mesoscale}
I.O. G{\"o}tze and G.~Gompper.
\newblock Mesoscale simulations of hydrodynamic squirmer interactions.
\newblock \emph{Physical Review E} 82, 041921 (2010).
\bibAnnoteFile{gotze2010mesoscale}

\bibitem{zhu2012self}
L.~Zhu, E.~Lauga, and L.~Brandt.
\newblock Self-propulsion in viscoelastic fluids: Pushers vs. pullers.
\newblock \emph{Physics of Fluids} 24, 051902 (2012).
\bibAnnoteFile{zhu2012self}

\bibitem{uspal2015rheotaxis}
W.E. Uspal, M.N. Popescu, S.~Dietrich, and M.~Tasinkevych.
\newblock Rheotaxis of spherical active particles near a planar wall.
\newblock \emph{Soft Matter} 11, 6613--6632 (2015).
\bibAnnoteFile{uspal2015rheotaxis}

\bibitem{wang2012inertial}
S.~Wang and A.~Ardekani.
\newblock Inertial squirmer.
\newblock \emph{Physics of Fluids} 24, 101902 (2012).
\bibAnnoteFile{wang2012inertial}

\bibitem{drescher2010direct}
K.~Drescher, R.~E. Goldstein, N.~Michel, M.~Polin, and I.~Tuval.
\newblock Direct measurement of the flow field around swimming microorganisms.
\newblock \emph{Physical Review Letters} 105, 168101 (2010).
\bibAnnoteFile{drescher2010direct}

\bibitem{downton2009simulation}
M.~T. Downton and H.~Stark.
\newblock Simulation of a model microswimmer.
\newblock \emph{Journal of Physics: Condensed Matter} 21, 204101 (2009).
\bibAnnoteFile{downton2009simulation}

\bibitem{pedley2016squirmers}
T.~J. Pedley, D.~R. Brumley, and R.~E. Goldstein.
\newblock Squirmers with swirl: a model for Volvox swimming.
\newblock \emph{Journal of Fluid Mechanics} 798, 165--186 (2016).
\bibAnnoteFile{pedley2016squirmers}

\bibitem{sonneborn1970methods}
T.M. Sonneborn.
\newblock Methods in Paramecium research.
\newblock In \emph{Methods in cell biology}, volume~4, pp. 241--339. Elsevier
  (1970).
\bibAnnoteFile{sonneborn1970methods}

\bibitem{zhang2015paramecia}
P.~Zhang, S.~Jana, M.~Giarra, PP. Vlachos, and S.~Jung.
\newblock Paramecia swimming in viscous flow.
\newblock \emph{The European Physical Journal Special Topics} 224, 3199--3210
  (2015).
\bibAnnoteFile{zhang2015paramecia}

\bibitem{ishikawa2006interaction}
T.~Ishikawa and M.~Hota.
\newblock Interaction of two swimming Paramecia.
\newblock \emph{Journal of Experimental Biology} 209, 4452--4463 (2006).
\bibAnnoteFile{ishikawa2006interaction}

\bibitem{ismagilov2002autonomous}
R.E. Ismagilov, A.~Schwartz, N.~Bowden, and G.M. Whitesides.
\newblock Autonomous movement and self-assembly.
\newblock \emph{Angewandte Chemie} 114, 674--676 (2002).
\bibAnnoteFile{ismagilov2002autonomous}

\bibitem{ozin2005dream}
G.A. Ozin, I.~Manners, S.~Fournier-Bidoz, and A.~Arsenault.
\newblock Dream nanomachines.
\newblock \emph{Advanced Materials} 17, 3011--3018 (2005).
\bibAnnoteFile{ozin2005dream}

\bibitem{Faivre2017}
P.~J. Vach, D.~Walker, P.~Fischer, P.~Fratzl, and D.~Faivre.
\newblock Pattern formation and collective effects in populations of magnetic
  microswimmers.
\newblock \emph{Journal of Physics D} 50, 11LT03 (2017).
\bibAnnoteFile{Faivre2017}

\bibitem{ren2017rheotaxis}
L.~Ren, D.~Zhou, Z.~Mao, P.~Xu, T.J. Huang, and T.E. Mallouk.
\newblock Rheotaxis of Bimetallic Micromotors Driven by Chemical--Acoustic
  Hybrid Power.
\newblock \emph{American Chemical Society Nano} 11, 10591--10598 (2017).
\bibAnnoteFile{ren2017rheotaxis}

\bibitem{golestanian2007designing}
R.~Golestanian, T.B. Liverpool, and A.~Ajdari.
\newblock Designing phoretic micro-and nano-swimmers.
\newblock \emph{New Journal of Physics} 9, 126 (2007).
\bibAnnoteFile{golestanian2007designing}

\bibitem{ebbens2010pursuit}
S.~J. Ebbens and J.~R. Howse.
\newblock In pursuit of propulsion at the nanoscale.
\newblock \emph{Soft Matter} 6, 726--738 (2010).
\bibAnnoteFile{ebbens2010pursuit}

\bibitem{sundararajan2008catalytic}
S.~Sundararajan, P.~E. Lammert, A.~W. Zudans, V.~H. Crespi, and A.~Sen.
\newblock Catalytic motors for transport of colloidal cargo.
\newblock \emph{Nano letters} 8, 1271--1276 (2008).
\bibAnnoteFile{sundararajan2008catalytic}

\bibitem{soler2014catalytic}
L.~Soler and S.~S{\'a}nchez.
\newblock Catalytic nanomotors for environmental monitoring and water
  remediation.
\newblock \emph{Nanoscale} 6, 7175--7182 (2014).
\bibAnnoteFile{soler2014catalytic}

\bibitem{gao2014environmental}
W.~Gao and J.~Wang.
\newblock The environmental impact of micro/nanomachines: a review.
\newblock \emph{American Chemical Society Nano} 8, 3170--3180 (2014).
\bibAnnoteFile{gao2014environmental}

\bibitem{Popescu2018}
M.~N. Popescu, W.~E. Uspal, Z.~Eskandari, M.~Tasinkevych, and S.~Dietrich.
\newblock Effective squirmer models for self-phoretic chemically active
  spherical colloids.
\newblock \emph{European Physical Journal E} 41, 145 (2018).
\bibAnnoteFile{Popescu2018}

\bibitem{Paxton2004}
W.F. Paxton, K.C. Kistler, C.C. Olmeda, A.~Sen, S.K.St. Angelo, Y.Y. Cao, T.E.
  Mallouk, P.E. Lammert, and V.H. Crespi.
\newblock Catalytic nanomotors: Autonomous movement of striped nanorods.
\newblock \emph{Journal of the American Chemical Society} 126, 13424 (2004).
\bibAnnoteFile{Paxton2004}

\bibitem{Paxton2006}
W.~F. Paxton, S.~Sundararajan, T.~E. Mallouk, and A.~Sen.
\newblock Chemical locomotion.
\newblock \emph{Angewandte Chemie International Edition} 45, 5420--5429 (2006).
\bibAnnoteFile{Paxton2006}

\bibitem{mathijssen2018oscillatory}
A.~Mathijssen, N.~Figueroa-Morale, G.~Junot, E.~Clement, A.~Lindner, and
  A.~Z{\"o}ttl.
\newblock Oscillatory surface rheotaxis of swimming E. coli bacteria.
\newblock \emph{Nature Communications} 10, 3434 (2019).
\bibAnnoteFile{mathijssen2018oscillatory}

\bibitem{uspal2013engineering}
W.~E. Uspal, H.~B. Eral, and P.~S. Doyle.
\newblock Engineering particle trajectories in microfluidic flows using
  particle shape.
\newblock \emph{Nature Communications} 4, 2666 (2013).
\bibAnnoteFile{uspal2013engineering}

\bibitem{jeffery1922motion}
G.B. Jeffery.
\newblock The motion of ellipsoidal particles immersed in a viscous fluid.
\newblock \emph{Proceedings of the Royal Society London A} 102, 161--179
  (1922).
\bibAnnoteFile{jeffery1922motion}

\bibitem{lettinga2005flow}
M.~P. Lettinga, Z.~Dogic, H.~Wang, and J.~Vermant.
\newblock Flow behavior of colloidal rodlike viruses in the nematic phase.
\newblock \emph{Langmuir} 21, 8048--8057 (2005).
\bibAnnoteFile{lettinga2005flow}

\bibitem{park2007cross}
J.~Park, J.~M. Bricker, and J.~E. Butler.
\newblock Cross-stream migration in dilute solutions of rigid polymers
  undergoing rectilinear flow near a wall.
\newblock \emph{Physical Review E} 76, 040801 (2007).
\bibAnnoteFile{park2007cross}

\bibitem{Kessler2004}
C.~Dombrowski, L.~Cisneros, S.~Chatkaew, R.~E. Goldstein, and J.~O. Kessler.
\newblock Self-concentration and large-scale coherence in bacterial dynamics.
\newblock \emph{Physical Review Letters} 93, 098103 (2004).
\bibAnnoteFile{Kessler2004}

\bibitem{Wensink2012}
H.~H. Wensink, J.~Dunkel, S.~Heidenreich, K.~Drescher, R.~E. Goldstein,
  H.~L{\"o}wen, and J.~M. Yeomans.
\newblock Meso-scale turbulence in living fluids.
\newblock \emph{Proceedings of the National Academy of Sciences of the United
  States of America} 109, 14308--14313 (2012).
\bibAnnoteFile{Wensink2012}

\bibitem{Goldstein2012}
F.~G. Woodhouse and R.~E. Goldstein.
\newblock Spontaneous circulation of confined active suspensions.
\newblock \emph{Physical Review Letters} 109, 168105 (2012).
\bibAnnoteFile{Goldstein2012}

\bibitem{Frey2018}
L.~Huber, R.~Suzuki, T.~Kr{\"u}ger, E.~Frey, and A.~R. Bausch.
\newblock Emergence of coexisting ordered states in active matter systems.
\newblock \emph{Science} 361, 255 -- 258 (2018).
\bibAnnoteFile{Frey2018}

\bibitem{Dogic2017}
K.-T. Wu, J.~B. Hishamunda, D.~T.~N. Chen, S.~J. DeCamp, Y.-W. Chang,
  A.~Fern{\'a}ndez-Nieves, S.~Fraden, and Z.~Dogic.
\newblock Transition from turbulent to coherent flows in confined
  three-dimensional active fluids.
\newblock \emph{Science} 355 (2017).
\bibAnnoteFile{Dogic2017}

\bibitem{Goldstein2016}
H.~Wioland, E.~Lushi, and R.~E. Goldstein.
\newblock Directed collective motion of bacteria under channel confinement.
\newblock \emph{New Journal of Physics} 18, 075002 (2016).
\bibAnnoteFile{Goldstein2016}

\bibitem{Clement2015}
H.~M. L\'opez, J.~Gachelin, C.~Douarche, H.~Auradou, and E.~Cl\'ement.
\newblock Turning bacteria suspensions into superfluids.
\newblock \emph{Physical Review Letters} 115, 028301 (2015).
\bibAnnoteFile{Clement2015}

\bibitem{DiLeonardo2017}
S.~Bianchi, F.~Saglimbeni, and R.~Di~Leonardo.
\newblock Holographic imaging reveals the mechanism of wall entrapment in
  swimming bacteria.
\newblock \emph{Physical Review X} 7, 011010 (2017).
\bibAnnoteFile{DiLeonardo2017}

\bibitem{DiLeonardo2018}
G.~Frangipane, D.~Dell'Arciprete, S.~Petracchini, C.~Maggi, F.~Saglimbeni,
  S.~Bianchi, G.~Vizsnyiczai, M.L. Bernardini, and R.~Di~Leonardo.
\newblock Dynamic density shaping of photokinetic E. coli.
\newblock \emph{eLife} 7, e36608 (2018).
\bibAnnoteFile{DiLeonardo2018}

\bibitem{Poon2018}
J.~Arlt, V.~A. Martinez, A.~Dawson, T.~Pilizota, and W.~C.~K. Poon.
\newblock Painting with light-powered bacteria.
\newblock \emph{Nature Communications} 9, 768 (2018).
\bibAnnoteFile{Poon2018}

\bibitem{Powers2009}
E.~Lauga and T.R. Powers.
\newblock The hydrodynamics of swimming microorganisms.
\newblock \emph{Reports on Progress in Physics} 72, 096601 (2009).
\bibAnnoteFile{Powers2009}

\bibitem{Ebbens2010}
S.J. Ebbens and J.R. Howse.
\newblock In pursuit of propulsion at the nanoscale.
\newblock \emph{Soft Matter} 6, 726--738 (2010).
\bibAnnoteFile{Ebbens2010}

\bibitem{Gompper2015_rev}
J.~Elgeti, R.G. Winkler, and G.~Gompper.
\newblock Physics of microswimmers -- single particle motion and collective
  behavior: a review.
\newblock \emph{Reports on Progress in Physics} 78, 056601 (2015).
\bibAnnoteFile{Gompper2015_rev}

\bibitem{Sen2010_rev}
Y.~Hong, D.~Velegol, N.~Chaturvedi, and A.~Sen.
\newblock Biomimetic behavior of synthetic particles: From microscopic
  randomness to macroscopic control.
\newblock \emph{Physical Chemistry Chemical Physics} 12, 1423 (2010).
\bibAnnoteFile{Sen2010_rev}

\bibitem{Sagues_review_2018}
A.~Doostmohammadi, J.~Ign{\'e}s-Mullol, J.~M. Yeomans, and F.~Sagu{\'e}s.
\newblock Active nematics.
\newblock \emph{Nature Communications} 9, 3246 (2018).
\bibAnnoteFile{Sagues_review_2018}

\bibitem{Shelley2013}
D.~Saintillan and M.~J. Shelley.
\newblock \emph{Comptes Rendus Physique} 14, 497 -- 517 (2013).
\bibAnnoteFile{Shelley2013}

\bibitem{Ramaswamy2002}
R.~Aditi~Simha and S.~Ramaswamy.
\newblock Hydrodynamic fluctuations and instabilities in ordered suspensions of
  self-propelled particles.
\newblock \emph{Physical Review Letters} 89, 058101 (2002).
\bibAnnoteFile{Ramaswamy2002}

\bibitem{Bechinger2016_rev}
C.~Bechinger, R.~Di~Leonardo, H.~L\"owen, C.~Reichhardt, G.~Volpe, and
  G.~Volpe.
\newblock Active particles in complex and crowded environments.
\newblock \emph{Reviews of Modern Physics} 88, 045006 (2016).
\bibAnnoteFile{Bechinger2016_rev}

\bibitem{felderhof2016stokesian}
B.U. Felderhof.
\newblock Stokesian swimming of a prolate spheroid at low Reynolds number.
\newblock \emph{European Journal of Mechanics-B/Fluids} 60, 230--236 (2016).
\bibAnnoteFile{felderhof2016stokesian}

\bibitem{leshansky2007frictionless}
A.M. Leshansky, O.~Kenneth, O.~Gat, and J.E. Avron.
\newblock A frictionless microswimmer.
\newblock \emph{New Journal of Physics} 9, 145 (2007).
\bibAnnoteFile{leshansky2007frictionless}

\bibitem{dassios1994generalized}
G.~Dassios, M.~Hadjinicolaou, and A.C. Payatakes.
\newblock Generalized eigenfunctions and complete semiseparable solutions for
  Stokes flow in spheroidal coordinates.
\newblock \emph{Quarterly of Applied Mathematics} 52, 157--191 (1994).
\bibAnnoteFile{dassios1994generalized}

\bibitem{lauga2016stresslets}
E.~Lauga and S.~Michelin.
\newblock Stresslets induced by active swimmers.
\newblock \emph{Physical Review Letters} 117, 148001 (2016).
\bibAnnoteFile{lauga2016stresslets}

\bibitem{theers2016modeling}
M.~Theers, E.~Westphal, G.~Gompper, and R.G. Winkler.
\newblock Modeling a spheroidal microswimmer and cooperative swimming in a
  narrow slit.
\newblock \emph{Soft Matter} 12, 7372--7385 (2016).
\bibAnnoteFile{theers2016modeling}

\bibitem{Michelin2014}
S.~Michelin and E.~Lauga.
\newblock Phoretic self-propulsion at finite {Pecl\'et} numbers.
\newblock \emph{Journal of Fluid Mechanics} 747, 572--604 (2014).
\bibAnnoteFile{Michelin2014}

\bibitem{ishimoto2017guidance}
K.~Ishimoto.
\newblock Guidance of microswimmers by wall and flow: Thigmotaxis and rheotaxis
  of unsteady squirmers in two and three dimensions.
\newblock \emph{Physical Review E} 96, 043103 (2017).
\bibAnnoteFile{ishimoto2017guidance}

\bibitem{katuri2018cross}
J.Katuri, W.E. Uspal, J.~Simmchen, A.~Miguel-L{\'o}pez, and S.~S{\'a}nchez.
\newblock Cross-stream migration of active particles.
\newblock \emph{Science Advances} 4, eaao1755 (2018).
\bibAnnoteFile{katuri2018cross}

\bibitem{Simmchen2016}
J.~Simmchen, J.~Katuri, W.E. Uspal, M.N. Popescu, M.~Tasinkevych, and
  S.~S{\'a}nchez.
\newblock Topographical pathways guide chemical microswimmers.
\newblock \emph{Nature Communications} 7, 10598 (2016).
\bibAnnoteFile{Simmchen2016}

\bibitem{happel2012low}
J.~Happel and H.~Brenner.
\newblock \emph{Low Reynolds number hydrodynamics: with special applications to
  particulate media}, volume~1.
\newblock Springer Science \& Business Media (2012).
\bibAnnoteFile{happel2012low}

\bibitem{popescu2010phoretic}
M.N. Popescu, S.~Dietrich, M.~Tasinkevych, and J.~Ralston.
\newblock Phoretic motion of spheroidal particles due to self-generated solute
  gradients.
\newblock \emph{The European Physical Journal E} 31, 351--367 (2010).
\bibAnnoteFile{popescu2010phoretic}

\bibitem{Abramowitz}
M.~Abramowitz and I.~A. Stegun.
\newblock \emph{Handbook of mathematical functions: with formulas, graphs, and
  mathematical tables}.
\newblock Dover (1970).
\bibAnnoteFile{Abramowitz}

\bibitem{ishikawa2008coherent}
T.~Ishikawa and T.J. Pedley.
\newblock Coherent structures in monolayers of swimming particles.
\newblock \emph{Physical Review Letters} 100, 088103 (2008).
\bibAnnoteFile{ishikawa2008coherent}

\bibitem{BEM}
C.~Pozrikidis.
\newblock \emph{A practical guide to boundary element methods with the software
  library BEMLIB}.
\newblock CRC Press (2002).
\bibAnnoteFile{BEM}

\bibitem{Howse2007}
J.~R. Howse, R.~A.~L. Jones, A.~J. Ryan, T.~Gough, R.~Vafabakhsh, and
  R.~Golestanian.
\newblock Self-motile colloidal particles: From directed propulsion to random
  walk.
\newblock \emph{Physical Review Letters} 99, 048102 (2007).
\bibAnnoteFile{Howse2007}

\bibitem{Baraban2012}
L.~Baraban, M.~Tasinkevych, M.~N. Popescu, S.~S{\'a}nchez, S.~Dietrich, and
  O.~G. Schmidt.
\newblock Transport of cargo by catalytic Janus micro-motors.
\newblock \emph{Soft Matter} 8, 48 (2012).
\bibAnnoteFile{Baraban2012}

\bibitem{anderson1989colloid}
J.L. Anderson.
\newblock Colloid transport by interfacial forces.
\newblock \emph{Annual Review of Fluid Mechanics} 21, 61--99 (1989).
\bibAnnoteFile{anderson1989colloid}

\bibitem{Golestanian2005}
R.~Golestanian, T.~B. Liverpool, and A.~Ajdari.
\newblock Propulsion of a molecular machine by asymmetric distribution of
  reaction products.
\newblock \emph{Physical Review Letters} 94, 220801 (2005).
\bibAnnoteFile{Golestanian2005}

\bibitem{Derjaguin1966}
B.V. Derjaguin, Yu.I. Yalamov, and A.I. Storozhilova.
\newblock Diffusiophoresis of large aerosol particles.
\newblock \emph{Journal of Colloid and Interface Science} 22, 117 -- 125
  (1966).
\bibAnnoteFile{Derjaguin1966}

\end{thebibliography}
\end{document}